\UseRawInputEncoding
\documentclass[conference,10pt]{IEEEtran}
\usepackage[english]{babel}
\usepackage{times}
\usepackage{multicol}
\usepackage{microtype}
\usepackage{marginnote}
\usepackage{framed}
\usepackage{roundbox}
\usepackage{setspace}
\usepackage{blindtext}
\usepackage{siunitx}
\usepackage{titlecaps}
\Addlcwords{or with if of and to -off off}

\usepackage{cite}

\usepackage{graphicx}
\usepackage{subfloat}
\usepackage{fancyhdr} 
\usepackage{color}
\usepackage{epsfig}
\graphicspath{{Images/}}
\usepackage{tikz}
\usepackage{tikzscale}
\usepackage{scalerel}
\usetikzlibrary{arrows}
\usetikzlibrary{plotmarks}
\usetikzlibrary{svg.path}
\usetikzlibrary{shapes.multipart}
\usepackage{pgfplots}
\pgfplotsset{compat=newest}
\pgfplotsset{every axis/.append style={
		label style={font=\Large},
		tick label style={font=\large}  
}}
\tikzstyle{int}=[draw, fill=black!10, minimum size=5em,thick]
\tikzstyle{init} = [pin edge={to-,thick,black}]

\usepackage{array}
\usepackage{stfloats}

\usepackage{amsthm,amsmath,amssymb,amsfonts}
\usepackage{relsize}
\usepackage{nicefrac}
\usepackage{bbm}
\usepackage{mathtools}
\usepackage[mathscr]{eucal}

\usepackage[ruled]{algorithm}
\usepackage{algpseudocode}
\usepackage{multirow}
\makeatletter
\gdef\Shortstack{\@ifnextchar[\@Shortstack{\@Shortstack[c]}}
\gdef\@Shortstack[#1]#2{%
	\leavevmode
	\vbox\bgroup
	\baselineskip-\p@\lineskip 3\p@
	\let\mb@l\hss\let\mb@r\hss
	\expandafter\let\csname mb@#1\endcsname\relax
	\let\\\@stackcr\setlength{\baselineskip}{#2}%
	\@ishortstack}
\makeatother
\usepackage{booktabs}
\usepackage{tabularx}

\usepackage{float}

\usepackage{url}
\usepackage[]{hyperref}
\usepackage[capitalize]{cleveref}

\newcommand\orcidicon[1]{\href{https://orcid.org/#1}{\includegraphics[scale=0.05]{orcid}}}

\newcommand{\myParagraph}[1]{{\bf \titlecap{#1}.}}

\newcommand{\tradeoff}{delay-accuracy trade-off\xspace}

\newcommand{\compDel}{processing time\xspace}
\newcommand{\commDel}{communication delay\xspace}

\DeclareMathOperator*{\argmax}{arg\,max}
\DeclareMathOperator*{\argmin}{arg\,min}

\newcommand{\sensSet}{\mathcal{V}}

\theoremstyle{plain}
\newtheorem{thm}{Theorem}

\newtheorem{lemma}{Lemma}
\theoremstyle{definition}
\newtheorem{definition}{Definition}
\newtheorem{prob}{Problem}
\newtheorem{ass}{Assumption}
\theoremstyle{remark}
\newtheorem{rem}{Remark}

\newcommand{\delayComp}{\tau}

\newcommand{\delayComm}{r(\delayComp)}

\algdef{SE}[DOWHILE]{Do}{doWhile}{\algorithmicrepeat}[1]{\algorithmicuntil\ #1}%

\renewcommand{\algorithmiccomment}[1]{\bgroup\hfill//~#1\egroup}

\addto\captionsenglish{\renewcommand{\figurename}{Fig.}}
\addto\captionsenglish{\renewcommand{\tablename}{TABLE}}

\newcommand{\blue}[1]{{\color{blue}#1}}

\newcommand{\linkToPdf}[1]{\href{#1}{\blue{(pdf)}}}
\newcommand{\linkToPpt}[1]{\href{#1}{\blue{(ppt)}}}
\newcommand{\linkToCode}[1]{\href{#1}{\blue{(code)}}}
\newcommand{\linkToWeb}[1]{\href{#1}{\blue{(web)}}}
\newcommand{\linkToVideo}[1]{\href{#1}{\blue{(video)}}}
\newcommand{\linkToMedia}[1]{\href{#1}{\blue{(media)}}}
\newcommand{\award}[1]{\xspace} 
\newcommand{\eg}{\emph{e.g.,}\xspace}
\newcommand{\ie}{\emph{i.e.,}\xspace}
\addto\extrasenglish{}
\addto\extrasenglish{}
\addto\extrasenglish{}

\title{
 \huge Computation and Communication Co-Design for Real-Time Monitoring and Control in Multi-Agent Systems}

\author{\authorblockN{Vishrant Tripathi\authorrefmark{1}, Luca Ballotta\authorrefmark{2}, Luca Carlone\authorrefmark{1}, and Eytan Modiano\authorrefmark{1}
\thanks{
The first two authors contributed equally to this paper.\newline
\indent This work was funded by NSF Grant CNS-1713725, by Army Research Office (ARO) grant no. W911NF-17-1-0508, by the Office of Naval Research under the ONR RAIDER program (N00014-18-1-2828),
by the CARIPARO Foundation Visiting Programme “HiPeR”,
and by the Italian Ministry of Education, University and Research
through the PRIN project no. 2017NS9FEY
and through the initiative ``Departments of Excellence" (Law 232/2016).
}
}
\authorblockA{\authorrefmark{1}Laboratory for Information and Decision Systems, MIT}
\authorblockA{\authorrefmark{2}Department of Information Engineering, University of Padova}
}

\IEEEoverridecommandlockouts

\begin{document}
\maketitle


\begin{abstract}
We investigate the problem of co-designing computation and communication in a multi-agent system (\eg a sensor network or a multi-robot team). 
We consider the realistic setting where each agent acquires sensor data and is capable of local processing before sending updates to a base station,
which is in charge of making decisions or monitoring phenomena of interest in real time. 
Longer processing at an agent leads to more informative updates but also larger delays,
giving rise to a \emph{\tradeoff} in choosing the right amount of local processing at each agent. 
We assume that the available communication resources are limited due to interference, bandwidth, and power constraints.
Thus, a scheduling policy needs to be designed to suitably share the communication channel among the agents.
To that end, we develop a general formulation 
to jointly optimize the local processing at the agents and the scheduling of transmissions. Our novel formulation {leverages} the notion of \emph{Age of Information} to quantify the freshness of data and capture the delays
caused by computation and communication. We develop efficient resource allocation algorithms using the Whittle index approach and
 demonstrate 
our proposed algorithms in two practical applications:
 multi-agent occupancy grid mapping in time-varying environments, and 
ride sharing in autonomous vehicle networks. Our experiments show that the proposed co-design approach  
leads to a substantial performance improvement (\boldmath$ 18-82\%$ in our tests). 

	
	\IEEEkeywords wireless networks; Age of Information; distributed computing; robotics; networked control systems.
\end{abstract}

\section{Introduction}

Monitoring and {control} of dynamical systems are fundamental and well-studied problems.
Many emerging applications involve performing these tasks over communication networks.
Examples include: sensing for IoT applications, control of robot swarms, real-time surveillance, and environmental monitoring by sensor networks.
Such systems typically involve multiple agents collecting and sending information to a central entity where data is stored, aggregated, analyzed,
and then used to send back control commands.
Due to the dramatic improvements both in on-device and edge computing,
and in wireless communication over the past two decades,
there has been a rapid growth in the size and scale of such networked systems. 

\begin{figure}[t]
    \centering
    \includegraphics[width=0.99\linewidth]{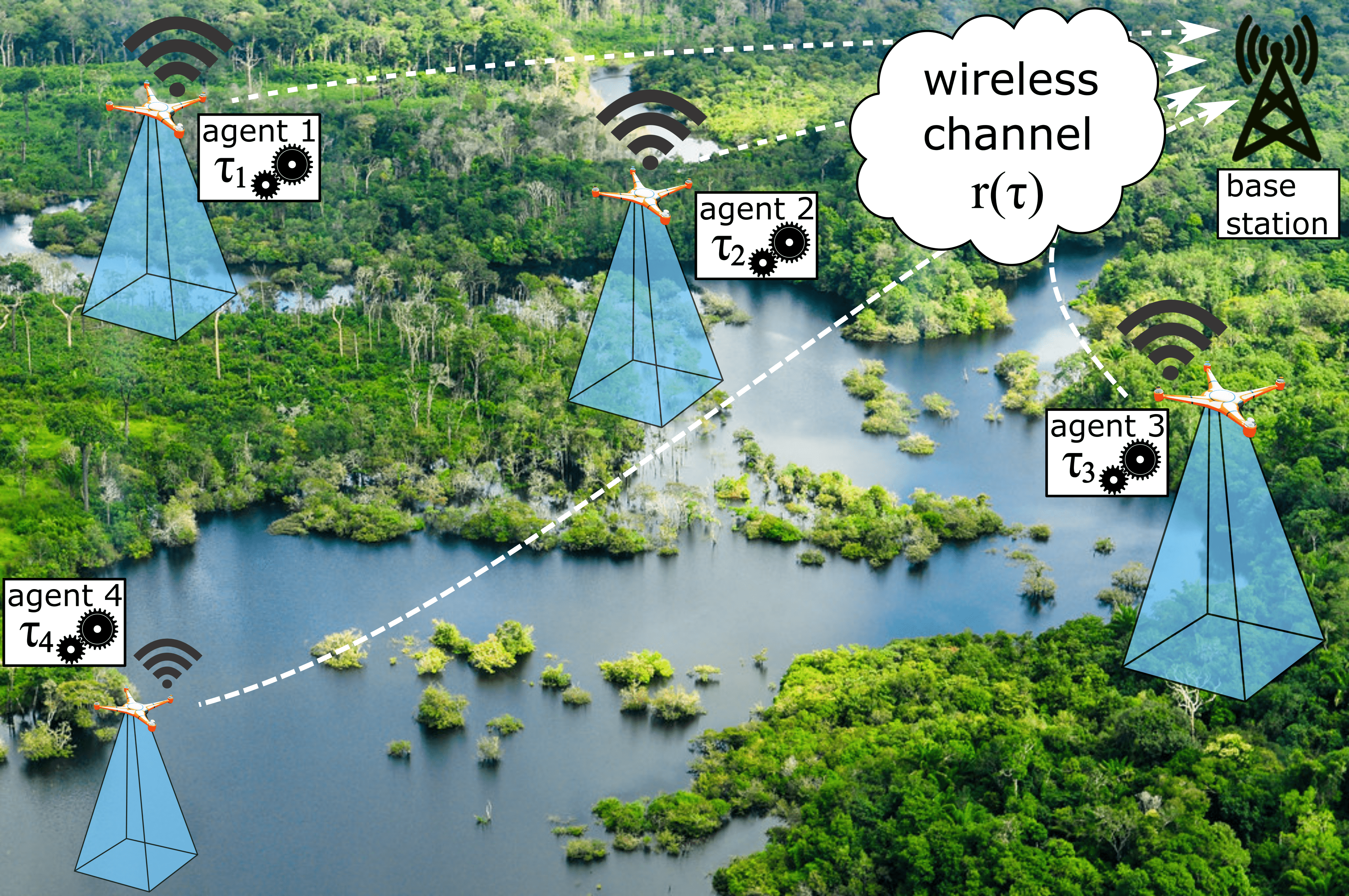}
    \vspace{-7mm}
    \caption{Example: four drones monitor different regions and send updates to a base station over a wireless channel.
    	Each agent spends time $\delayComp_i$ processing the collected measurements before sending.
    	A scheduling algorithm prioritizes transmissions to the base station.
    	This paper focuses on the co-design of the processing times $\delayComp_i$ and the scheduling policy.}  
    \label{fig:example}
    \vspace{-6mm}
\end{figure}

This has motivated the design of scalable architectures, both for computation and communication.
Two key directions of innovation involve
a) pushing the computation to be distributed across the network,
such that all agents perform local processing of the collected measurements,
and b) designing scheduling algorithms that efficiently share limited communication resources across all devices and ensure timely delivery of information.
However, existing work on communication scheduling~\cite{kaul2012real,yin17_tit_update_or_wait,kadota2018scheduling,talak2018optimizing,tripathi2019whittle} disregards distributed processing, 
while related work on sensor fusion~\cite{xiao2005scheme,carli2008distributed,olfati2005consensus} focuses on designing distributed algorithms, 
rather than allocating computational resources at each node.

In this work, we explore the joint optimization of computation and communication resources for  monitoring and control tasks.
We consider a multi-agent system
where {each agent is in charge of monitoring a time-varying phenomenon} and sending information to a central base station. For instance, this setup can model a team of robots mapping a dynamic environment and sending map updates to a base station, which aggregates a global map for centralized decision-making (\autoref{fig:example}). 

The agents are capable of local processing before transmitting the acquired information. 
This could involve operations such as refining, denoising, or compressing the data or simply gathering more informative updates.
We assume that the more time an agent spends in processing locally,
the higher the quality of the generated update.
However, longer processing also induces a delay in between subsequent updates.
This yields a \emph{\tradeoff}:
is it better to send outdated but high-quality updates, 
or to reduce the overall latency by communicating low-quality information?

We consider the realistic scenario where the total communication resources available are limited
due to interference, limited bandwidth, and/or power constraints. Thus, in any given time-slot, only one of the agents is allowed to communicate with the base station.
The communication constraints mean that,
in addition to optimizing the local processing times,
a \textit{scheduling policy} needs to be designed to specify which agents can communicate in every time-slot.

Therefore, the goal of this work is to develop a general framework to determine
 the optimal amount of local processing at each agent in the network and 
design a scheduling policy to prioritize communication in order to maximize performance.

\myParagraph{Related work}
Over the past few years, there has been a rapidly growing body of work using \emph{Age of Information} (AoI) as a metric for designing scheduling policies in communication networks~\cite{kadota2018scheduling,talak2018optimizing,tripathi2019whittle}
and for control-driven tasks in networked control systems~\cite{sun2017remote,ornee2019sampling,champati2019performance,klugel2019aoi}.
AoI captures the timeliness of received information at the destination (see \cite{kosta2017age,sun2019age_book} for recent surveys). 
Our processing and scheduling co-design problem is motivated by recent advances in embedded electronics,
as well as the development of efficient estimation and inference algorithms for real-time applications on low-powered devices~\cite{NNRuntime,sandler2018mobilenetv2, 2018arXiv180402767R, howard2019searching}.
The output accuracy of such algorithms increases with the runtime,
in line with the \tradeoff we consider in this paper.
Another application of such a trade-off involves deciding on computation offloading in cloud robotics,
which has been the focus of recent works on real-time inference by resource-constrained robots~\cite{crankshaw2017clipper, Pavone-RSS-19}. In this context, sending raw data can induce long transmission delays, but allow better inference by shifting the computational burden the cloud.

\myParagraph{Contributions} 
We address the computation and communication co-design problem and 
develop a) a scheduling policy that ensures timely delivery of updates, 
and b) an algorithm to determine the optimal amount of local processing at each agent.
To do so, we use AoI to 
measure the lag in obtaining information for monitoring and {control} of time-critical systems.
Our contribution is threefold.
First, we develop a general framework to jointly optimize computation and communication for real-time monitoring and decision-making (\autoref{sec:system-model}).
This framework extends existing work~\cite{ballotta2019computationcommunication} by
a) considering joint optimization of scheduling in addition to processing, 
and b) addressing a general model that goes beyond linear systems. 

Second, we develop low-complexity scheduling and processing allocation schemes that perform well in practice (Sections~\ref{sec:lagrangian}-\ref{sec:whittle}). The co-design problem is a multi-period resource allocation problem and is hard to solve in general due to its combinatorial nature. We resolve this by considering a Lagrangian relaxation that decouples the problem into multiple single-agent problems, which can be solved effectively. To solve the scheduling problem, we generalize the Whittle index framework proposed in \cite{tripathi2019whittle} for sources that generate updates at different rates and of different sizes.

Finally, we demonstrate the benefits of using our methods in two practical applications from robotics and autonomous systems: multi-agent occupancy grid mapping in time-varying environments, and ride-sharing systems with local route optimization (\autoref{sec:applications}). Our simulations show that we can achieve performance improvements of $18-35\%$ in the mapping application and $75-82\%$ in the ride-sharing application with respect to baseline approaches. 






\section{Problem Formulation}
\label{sec:system-model}

We consider a discrete-time setting with $N$ agents in a networked system, where each agent is in charge of monitoring a time-varying phenomenon and sending information updates to a base station. Each agent processes the collected measurements locally, before sending its updates. The $i$-th agent spends $\delayComp_i $ time slots to process a new update. We refer to this quantity as the \emph{processing time} associated with agent $i$.

We assume that sensing and processing happen sequentially at each agent.
Thus, agent $i$ acquires a new sample every $ \delayComp_i $ time slots. 
Further, each agent stores in a buffer the freshest processed measurement. 
We will assume that the processing time allocations $ \delayComp_i, \forall i $  are constant during operation. 

To communicate the acquired and processed updates, the agents use a wireless communication channel. We assume that, due to interference and bandwidth constraints, only one of the agents can transmit to the base station in any given time-slot. At every transmission opportunity, the base station polls one of the agents regarding the state of its system and receives the most recent measurement that has been processed. 





Scheduling decisions are modeled as indicator variables $ u_i(t) $
where $ u_i(t) = 1 $ if the $i$-th agent is scheduled at time $ t $
and zero otherwise. We assume that a transmission from the $ i $-th agent takes $ r_i(\delayComp_i) $ time slots,
with $ r_i(\cdot) $ a monotone sequence. This captures one aspect of the delay-accuracy trade-off, namely that the size of the update depends on the amount of time spent in processing it.
When the agents spend local processing to collect more detailed information, \eg in exploration tasks, the measurements get larger overtime and $ r_i(\cdot) $ is increasing. 
Conversely, when the agents compress the collected data, \eg extracting visual features from images, $ r_i(\cdot) $ is decreasing.



\begin{figure}
	\centering
	\includegraphics[width=0.99\linewidth]{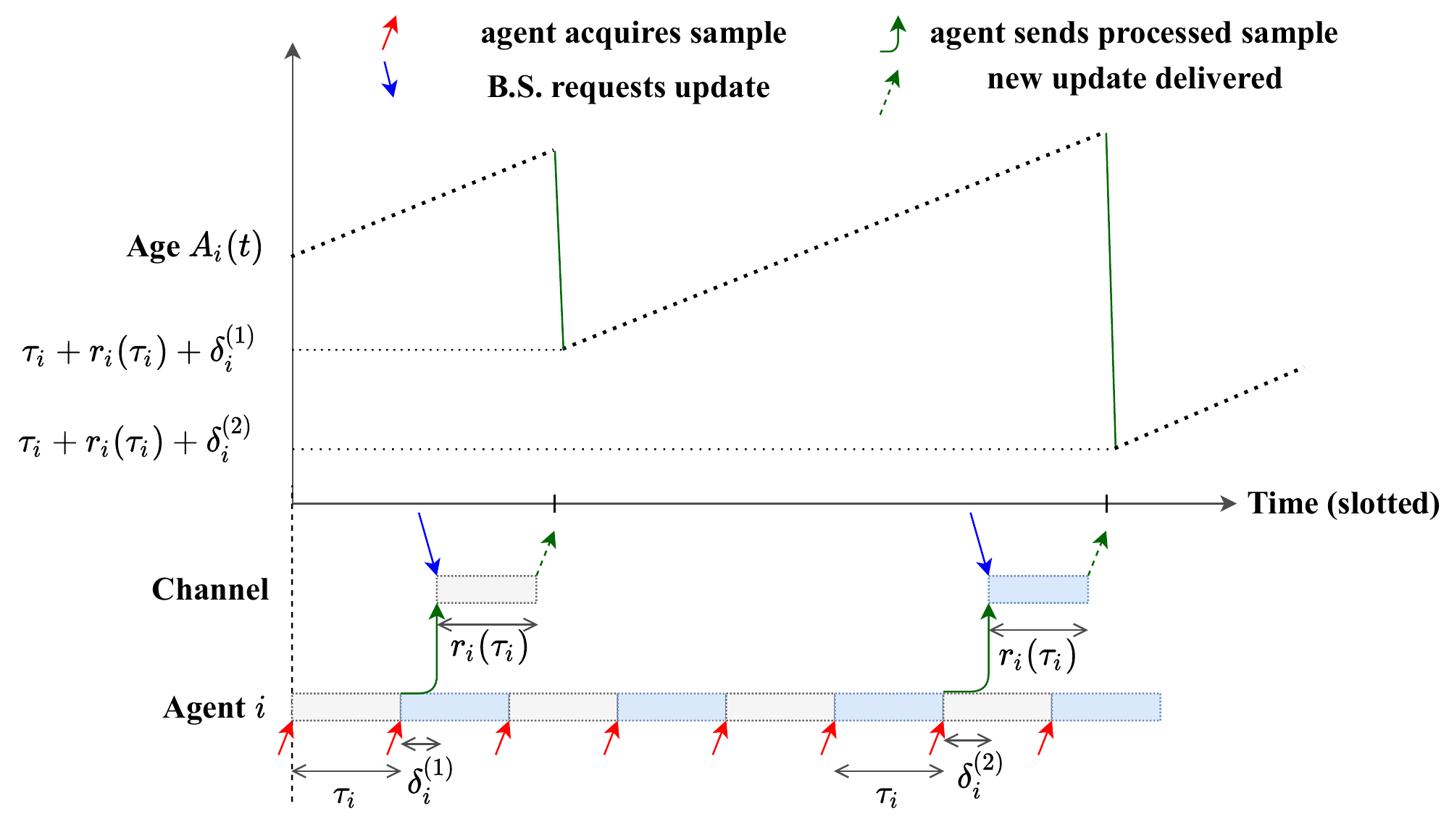}\vspace{-3mm}
	\caption{AoI evolution for agent $i$.
		The agent acquires and processes new samples every $\tau_i$ time-slots.
		When the base station (B.S.) requests a new update, the agent sends the most recent sample that has finished processing, taking $r_i(\tau_i)$ time-slots {for transmission}.
		The variable $\delta_i^{(k)}$ represents the waiting time in the buffer for update $k$.
		Upon a new update delivery, the AoI at the base station $A_i(t)$ drops to the age of the delivered update.
		\label{fig:singleSensor} \vspace{-5mm}}
\end{figure}

To measure the freshness of the information at the base station, we use a metric called Age of Information (AoI). The AoI $ A_i(t) $ measures how old the information at the base station is regarding agent $i$ at time $t$. Upon receiving a new update, it drops to the age of the delivered update. Otherwise, it increases linearly. The evolution is described below:
\begin{equation}
\label{eq:AoIEvolution}
A_{i}(t+1) =
\begin{cases}
\tau_i + r_i(\tau_i) + \delta_i^{(k)}, &\text{ if update } k \text{ is delivered},\\
A_{i}(t)+1, &\text{ otherwise.}
\end{cases}
\end{equation}
Here $\delta_i^{(k)}$ is the \emph{waiting time} spent by the $ k $-th update from agent $ i $ in the buffer, \ie the delay from the time the update was processed to the time it was actually transmitted. 
Since a new processed update is generated every $\tau_i$ time-slots, the waiting time $\delta_i^{(k)}$ ranges from $0$ to $\tau_i - 1$ time-slots.
\autoref{fig:singleSensor} depicts the AoI process for agent $i$. Observe that the lowest value that the AoI can drop to is $\tau_i + r_i(\tau_i)$, since every update spends time $\tau_i$ in processing and time $r_i(\tau_i)$ in communication.

The AoI evolution in~\eqref{eq:AoIEvolution} is involved since it requires analyzing waiting times that vary with each update.
To simplify the analysis,
while still capturing the relevant {features of the AoI dynamics}, 
we assume that 
the sequences $\delta_i^{(k)}$ are constant over time, \ie $\delta_i^{(k)}\equiv\delta_i \,\forall k, \, \forall i\in\sensSet$.
Each $\delta_i$ accounts for the average waiting time accumulated by a processed measurement before it is sent by the $i$-th agent. We are interested in the practical setting where processing times $\tau_i$ are small, and the number of agents $N$ is large.
Thus, our assumption of constant waiting times is reasonable, since the waiting time's contribution to the overall AoI is negligible on average (being upper bounded by $\tau_i$) 
as compared to the time between subsequent requests from the base station, which grows linearly with the number of agents $N$ \cite{kadota2018scheduling}. 
The smallest AoI for agent $ i $ is defined as $\Delta_i\triangleq\tau_i + r_i(\tau_i) + \delta_i $,
which is the value that AoI resets to upon a new update delivery.

It has been shown in recent works \cite{sun2017remote,ornee2019sampling,champati2019performance, klugel2019aoi} that real-time monitoring error for linear dynamical systems can be seen as an increasing function of the AoI. Intuitively, fresher updates lead to higher monitoring accuracy and better control performance.
%
%
Motivated by this, we assume that each agent has an associated cost function $J_i(\delayComp_i,A_i(t)) $ that maps the processing time and the current AoI to a cost that reflects how useful the current information at the base station is for monitoring or control.  

\begin{ass}[Delay-Accuracy Trade-off]\label{ass:trade-off}
The cost functions $J_i(\delayComp_i,A_i(t)) $ are 
increasing with the AoI $A_i(t)$
and decreasing with the processing time $ \delayComp_i $. Thus, 
longer processing leads to more useful measurements (for a fixed age), 
while fresher information 
induces a lower cost than outdated information. 
\end{ass}

\begin{rem}[Task-related cost function]\label{rem:task-related-cost}
	The functional form of $J_i(\delayComp_i,A_i(t))$ depends on the underlying dynamics of the system $i$ and 
	on the impact of agent processing on the quality of updates. These functions are typically estimated using domain knowledge or learned from data offline. 
	The approach in this paper holds for any functions $J_i(\delayComp_i,A_i(t))$ that satisfy the above assumption.	We discuss numerical examples in \autoref{sec:applications}.
\end{rem}


Our goal is to design a {causal} scheduling policy $\pi$ and find the processing times $ \tau_1,...,\tau_N $
for every agent so as to minimize the sum of the time-average costs.
\begin{framed}
\begin{prob}[Computation and Computation Co-design]\label{prob:codesign}
	Given the set of agents $ \sensSet =\{1,\dots,N\}$, cost functions $ \{J_i\left(\cdot,\cdot\right)\}_{i\in\sensSet} $,
	and AoI evolution~\eqref{eq:AoIEvolution},
	find the {processing times} $ \{\delayComp_i \}_{i\in\sensSet} $ and the scheduling policy $ \pi $ 
	that minimize the infinite-horizon time-averaged cost:
	\begin{align}\tag{P1}
    \begin{aligned}
    \label{eq:problem-statement}
    		&\min_{\substack{\delayComp_i \in \mathcal{T}_i\, \forall i\in\sensSet \\\pi\in\Pi}} &&
    		\sum_{i\in\sensSet}\limsup_{T\rightarrow+\infty}\mathbb{E}_\pi\left[\dfrac{1}{T}\sum_{t=t_0}^{T}J_i\left(\delayComp_i,A_i^\pi(t)\right)\right]\\
    		& \text{\hspace{5mm}s.t.} &&\sum_{i\in\sensSet} u_i^{\pi}(t) \le 1, \forall t
    \end{aligned}
    \end{align}
	where {$\Pi$ is the set of causal scheduling policies},
	$u_i^{\pi}(t) = 1$ if policy $\pi$ schedules agent $i$ at time $t$ and $u_i^{\pi}(t)=0$ otherwise.
	$\mathcal{T}_i$ is the set of admissible processing times for agent $i$, and $ A^\pi_i(t) $ is the AoI of the $ i $-th agent at time $ t $ under policy $ \pi $.
\end{prob}
\end{framed}
Finding the optimal processing times requires iterating over the combinatorial space $\mathcal{T}_i\times ... \times\mathcal{T}_N$, while finding the optimal scheduling policy 
requires solving a dynamic program which suffers from the curse of dimensionality.
\section{A Lagrangian Relaxation}
\label{sec:lagrangian}

We now discuss a relaxation of~\cref{prob:codesign} that enables us to develop efficient algorithms.
This approach is motivated by the work of Whittle~\cite{whittle1988restless} and its applications to network scheduling \cite{tripathi2019whittle}. 
The relaxation will be useful not only for finding a scheduling policy, but also in optimizing the processing times. 



We start by considering a relaxation of \eqref{eq:problem-statement} where the scheduling constraint is to be satisfied on average,
rather than at each time slot.
The relaxed problem is given by
\begin{align}
\begin{aligned}
\label{eq:problem-relax}
		&\min_{\substack{\delayComp_i \in \mathcal{T}_i\, \forall i\in\sensSet \\\pi\in\Pi}} &&
		\sum_{i\in\sensSet}\limsup_{T\rightarrow+\infty}\mathbb{E}_\pi\left[\dfrac{1}{T}\sum_{t=t_0}^{T}J_i\left(\delayComp_i,A_i^\pi(t)\right)\right]\\
		& \text{\hspace{5mm}s.t.} &&\sum_{i\in\sensSet} \limsup_{T\rightarrow+\infty} \frac{\sum_{t=t_0}^{T} u_i^\pi(t)}{T} \le 1.
\end{aligned}
\end{align}

To solve~\eqref{eq:problem-relax}, we introduce a Lagrange multiplier $C > 0$ for the average scheduling constraint.
The Lagrange optimization is given by the following equation:
\begin{gather}\label{eq:decoupled-problem-tau}
	\max_{C > 0}\min_{\substack{\tau_i \in \mathcal{T}_i\, \forall i\in\sensSet \\ \pi\in\Pi}}
	\;\;\sum_{i\in\sensSet}\bar{J}_i(\delayComp_i,C)- C\\
	\nonumber\bar{J}_i(\delayComp_i,C)\triangleq\limsup_{T\rightarrow+\infty}\mathbb{E}_\pi\left[\dfrac{1}{T}\sum_{t=t_0}^{T}\bigg(J_i\left(\delayComp_i,A_i^\pi(t)\right)+Cu^\pi_i(t)\bigg)\right] 
\end{gather}

Due to the Lagrangian relaxation,
the inner minimization can be decoupled as the sum of $ N $ independent problems. 
\begin{framed}
\begin{prob}[Decoupled Problem $i$]\label{prob:decoupled-problem}
	Given a constant cost $ C > 0 $, 
	find a scheduling policy $ \pi_i = \{u_i(t)\}_{t\ge t_0} $ and a processing time $\tau_i \in \mathcal{T}_i$ that minimize the infinite-horizon time-averaged cost of agent $ i $:
	\begin{equation}\label{eq:decoupled-problem}\tag{P2}
		{\min_{\substack{\delayComp_i \in \mathcal{T}_i \\\pi_i\in\Pi}}}\;\;
		\limsup_{T\rightarrow+\infty}\mathbb{E}_{\pi_i}\left[\dfrac{1}{T}\sum_{t=t_0}^{T}\bigg(J_i\left(\delayComp_i,A_i^{\pi_i}(t)\right)+Cu_i(t)\bigg)\right]
	\end{equation}
\end{prob}
\end{framed}
In~\cref{prob:decoupled-problem}, the multiplier $C$ can be interpreted as a transmission cost: whenever $u_i(t)=1$, agent $i$ has to pay a cost of $C$ for using the channel. Further, transmitting an entire update costs $C r_i(\tau_i)$, since $i$ transmits for $r_i(\tau_i)$ time-slots.\\
In the next section, we look at the single-agent problem~\eqref{eq:decoupled-problem} in greater detail, and show how to solve it exactly. Since the problem involves a single agent, it is much easier to solve than the original combinatorial formulation. The solution also provides key insights in choosing both the scheduling policy and the processing times for the original problem~\eqref{eq:problem-statement}.


\subsection{Solving the Decoupled Problem}\label{sec:decoupled-problem}
We now solve~\cref{prob:decoupled-problem} for each agent separately.
First, we characterize the structure of the optimal scheduling policy $\pi_i^*$ given a fixed value of $\tau_i$.
Then, we optimize over the latter.

\begin{thm}\label{thm:optimal-threshold}
	The solution to~\cref{prob:decoupled-problem},
	given a fixed value of $\tau_i$,
	is a stationary threshold-based policy:
	let $ \widetilde{H}_i \triangleq H_i + r_i(\tau_i)$ and suppose there exists an age $H_i$ that satisfies
	\begin{equation}\label{eq:optimal-threshold}
		J_i(\delayComp_i,\widetilde{H}_i - 1) \le J_i^W(\delayComp_i,H_i)	\le J_i\big(\delayComp_i,\widetilde{H}_i)
	\end{equation}
	\noindent where 
	\begin{equation}\label{eq:whittle-cost}
		J_i^W(\delayComp_i,H_i) \triangleq \dfrac{\sum_{h=\Delta_i}^{\widetilde{H}_i-1}J_i(\delayComp_i,h) + Cr_i(\tau_i)}{\widetilde{H}_i - \Delta_i}. 
	\end{equation}
	Then, an optimal scheduling policy $\pi_i^*$ is to start sending an update whenever $A_i(t) \geq H_i$ and to not transmit otherwise. If no such $H_i$ exists, the optimal policy is to never transmit. The quantity $J_i^W(\delayComp,H_i)$ represents the time-average cost of using a threshold policy with the AoI threshold $H_i$. 
\end{thm}
\begin{IEEEproof}
	See Appendix~\ref{app:threshold-policy}.
\end{IEEEproof}

The structure of the optimal scheduling policy $\pi_i^*$
according to~\cref{thm:optimal-threshold} is intuitive, due to the monotonicity of the cost functions $J_i(\tau_i,\cdot)$ in the AoI. If it is optimal to transmit and pay the cost $C$ for $r_i(\tau_i)$ time-slots at a particular AoI, it should be also be optimal to do so when the AoI is higher, since the gain from AoI reduction would be even more.
Given $\tau_i$ and $C$ , a way to compute the optimal threshold is to start from $H_i = \Delta_i$ and increase $H_i$ until condition \eqref{eq:optimal-threshold} is satisfied.
Let the value that this procedure terminates at be denoted by $ H_i(\delayComp_i) $. Then, $ H_i(\delayComp_i) $ is an optimal threshold for agent $i$.

Next, we look at how to compute the optimal processing time $ \delayComp_i^* $ to solve Problem \ref{prob:decoupled-problem}.
To do so, given the admissible set $ \mathcal{T}_i$, 
we find the value of $\tau_i\in\mathcal{T}_i$ that induces the lowest time-averaged cost
for agent $i$ {by enumerating over the set $ \mathcal{T}_i$}:
\begin{equation}\label{eq:optimal-tau-decoupled}
	\delayComp_i^* = \argmin_{\delayComp_i \in \mathcal{T}_i} \tilde{J}_i^W(\delayComp_i).
\end{equation}
where $ \tilde{J}_i^W(\delayComp_i) \triangleq J_i^W\big(\delayComp_i,H_i(\delayComp_i)\big) $.
The optimal processing times $\tau_i^{*}$ and
policies $\pi_i^*$, with thresholds $H_i(\tau_i^{*})$,
computed for each decoupled problem
provide an optimal solution to the inner minimization of \eqref{eq:decoupled-problem-tau}.

\subsection{Optimizing Processing Times in \cref{prob:codesign}}\label{sec:processing-optimization}

Leveraging the solution of the decoupled problems found in~\autoref{sec:decoupled-problem},
we now design a procedure to optimize the processing times for the original multi-agent~\cref{prob:codesign}. 

Given a cost $C > 0$, we can use \eqref{eq:optimal-threshold} and \eqref{eq:optimal-tau-decoupled} to compute the optimal processing times $\tau_i^*$ and the corresponding AoI thresholds $ H_i(\delayComp_i^*) $ for the $N$ decoupled problems in \eqref{eq:decoupled-problem-tau}. 
Further, observe that, for the $i$-th decoupled problem, the optimal scheduling policy for agent $i$ chooses to send a new update every time the AoI exceeds $H_i(\tau_i^*)$ and the AoI drops to $\Delta_i$ after each update delivery. 
Thus, the fraction of time that agent $i$ occupies the channel (on average) is given by
\begin{equation}\label{eq:sensor-frequency-decoupled}
	f_i(\delayComp_i^*) = \dfrac{r_i(\delayComp_i^*)}{H_i(\delayComp_i^*) + r_i(\tau_i^*) - \Delta_i}.
\end{equation}

The total channel utilization given the Lagrange multiplier $C$ is $ f = \sum_{i\in\sensSet}f_i(\delayComp_i^*) $.
{From~\eqref{eq:problem-relax}, $f$ must lie in the interval $[0,1]$ to represent a feasible allocation of computation and communication resources.
If not, then more than one agent is transmitting in every time-slot \textit{on average},
which is not possible given the (relaxed) interference constraint.}



This suggests a natural way to optimize over both the Lagrange cost $ C $
and the processing times $\tau_i$,
which is presented in~\cref{alg:optimal-processing}.
In particular, 
we optimize the processing times $\tau_i$ by using \eqref{eq:optimal-threshold} and \eqref{eq:optimal-tau-decoupled} (line~\ref{alg:optimization-tau} in Algorithm \ref{alg:optimal-processing}), and update $ C $ via a dual-ascent scheme (lines~\ref{alg:ascent-C-begin}--\ref{alg:ascent-C-end}) using the average channel utilization $f_{\textit{curr}}$.

\begin{algorithm}
	\caption{Optimizing Processing Times}
	\label{alg:optimal-processing}
	\begin{algorithmic}[1]
		\Require Costs $ J_i^W(\cdot) $, set of admissible processing times $ \mathcal{T}_i $ for each agent $ i\in\sensSet $, stepsize $ \alpha > 0 $. 
		\Ensure Locally optimal processing times $ \{\delayComp_i^*\}_{i\in\sensSet} $.
		
				\State $ C \leftarrow C_0 $;
				\Loop 
					\For{sensor $ i\in\sensSet $} \Comment{optimization~\eqref{eq:optimal-tau-decoupled}}
						\State $ \delayComp_i^* \gets \argmin_{\delayComp_i \in \mathcal{T}_i} \tilde{J}_i^W(\delayComp_i) $;
							\label{alg:optimization-tau}
					\EndFor
					\State $ f_{\textit{curr}} \gets \sum_{i\in\sensSet} f_i(\delayComp_i^*) $;\label{alg:ascent-C-begin}
					\State $ C \gets C + \alpha(f_{\textit{curr}}-1) $;\label{alg:ascent-C-end}
				\EndLoop
		\State \Return $ \{\delayComp_i^*\}_{i\in\sensSet} $.
	\end{algorithmic}
\end{algorithm}

Intuitively, the algorithm keeps increasing the virtual communication cost (quantified by the Lagrange multiplier $C$) until the processing times computed in line~\ref{alg:optimization-tau} become compatible with the scheduling constraint. 
The decoupling reduces the complexity of finding the optimal processing times
from combinatorial $O\left(\prod_{i \in \mathcal{V}} |\mathcal{T}_i|\right)$ to linear search $O\left(\sum_{i \in \mathcal{V}} |\mathcal{T}_i|\right)$.

\section{Whittle-index Scheduling}
\label{sec:whittle}



In the previous section, we established a threshold structure for the optimal scheduling policy of the relaxed problem~\eqref{eq:problem-relax},
where each agent transmits when its AoI exceeds $H_i(\tau_i^*)$. Next, we exploit this threshold structure to design an efficient scheduling policy for the original~\cref{prob:codesign}.
Given the processing times $\tau_i^*$ computed via~\cref{alg:optimal-processing},
we need to solve:
\begin{align}
\begin{aligned}
\label{eq:problem-scheduling}
		&\min_{\substack{\pi\in\Pi}} &&
		\sum_{i\in\sensSet}\limsup_{T\rightarrow+\infty}\mathbb{E}_\pi\left[\dfrac{1}{T}\sum_{t=t_0}^{T}J_i\left(\delayComp_i^*,A_i^\pi(t)\right)\right]\\
		& \text{\hspace{5mm}s.t.} &&\sum_{i\in\sensSet} u_i(t) \le 1, \forall t.
\end{aligned}
\end{align}

Minimizing the time-average of increasing functions of AoI was considered in~\cite{tripathi2019whittle}. 
There, the authors introduced a low-complexity near-optimal scheduling policy using the Whittle index approach. Unlike the setting in~\cite{tripathi2019whittle}, our agents generate updates at different rates (every $\tau_i$ time-slots for agent $i$) and
induce different communication delays ($r_i(\tau_i)$ time-slots).
We now generalize the Whittle index approach for our setting. 

The Whittle index approach consists of four steps:
1) converting the problem into an equivalent restless multi-armed bandit (RMAB) formulation, 2) decoupling the problem via a Lagrange relaxation, 3) establishing a structural property called \textit{indexability} for the decoupled problems, and 4) using this structure to formulate a Whittle index policy for the original scheduling problem. We go through these steps below.

\myParagraph{Step 1} We first need to establish \eqref{eq:problem-scheduling} can be equivalently formulated as a restless multi-armed bandit problem. We do so in Appendix~\ref{app:rmab}. 

\myParagraph{Step 2} As we observed in~\autoref{sec:lagrangian}, the original scheduling problem can be split into $N$ decoupled problems of the form \eqref{eq:decoupled-problem} via a Lagrange relaxation.
Further, through~\cref{thm:optimal-threshold}, we know that the optimal scheduling policy for each decoupled problem has a threshold structure, \ie agent $i$ should transmit only if its associated AoI $A_i(t)$ exceeds the threshold $H_i(\tau_i^*)$.

\myParagraph{Step 3} Whittle showed in~\cite{whittle1988restless} that when there is added structure in the form of a property called \textit{indexability} for the decoupled problems, then the RMAB admits a low-complexity solution called the Whittle index, that is known to be near optimal \cite{weber1990index}. The \textit{indexability} property for the $i$-th decoupled problem requires that, as the transmission cost $C$ increases from $0$ to $\infty$, the set of AoI values for which it is optimal for agent $i$ to transmit must decrease monotonically from the entire set (all ages $A_i(t)\ge \Delta_i$) to the empty set (never transmit). In other words, the optimal threshold $H_i(\tau_i^*)$ should increase as the transmission cost $C$ increases. Next, we use Theorem \ref{thm:optimal-threshold} and the monotonicity of the cost functions $J_i(\tau_i^*,\cdot)$ to establish that the decoupled problems are indeed indexable.
\begin{lemma}
\label{lem:indexability}
The \textit{indexability} property holds for the decoupled problems \eqref{prob:decoupled-problem}, given an allocation of processing times $\tau_i$.
\end{lemma}
\begin{IEEEproof}
	See Appendix~\ref{app:indexability}.
\end{IEEEproof}

\myParagraph{Step 4} Having established indexability for the decoupled~\cref{prob:decoupled-problem},
we can derive a functional form for the Whittle index which solves the scheduling for the original~\cref{prob:codesign}.
\begin{framed}
\begin{definition}\label{def:whittle-index}
	For the $i$-th decoupled problem, 
	the Whittle index $ W_i(H) $ is defined as the minimum cost $ C $
	that makes both scheduling decisions (transmit, not transmit) equally preferable at AoI $H$.
	Let $\widetilde{H} \triangleq H+r_i(\tau_i)$.
	The expression for $ W_i(H) $, given a processing time $\tau_i$, is: 
	\begin{equation}
	    \label{eq:WhittleIndex}
	    W_i(H) \triangleq \frac{\big(\widetilde{H}-\Delta_i\big)J\big(\tau_i,\widetilde{H}\big) - \sum\limits_{k=\Delta_i}^{\widetilde{H}-1} J(\delayComp_i,k) }{r_i(\tau_i)}.
	\end{equation}
\end{definition}
\end{framed}

We derive the expression above in~\cref{app:indexability}.
Using \eqref{eq:WhittleIndex}, we can now design the Whittle index policy to solve \eqref{eq:problem-scheduling}. Whenever the channel is unoccupied,
the agent with the most critical update should be asked for an update. This leads to the scheduling policy presented in~\cref{alg:optimal-scheduling}. The Whittle index policy chooses the agent with the highest index (line~\ref{alg:whittle-scheduling}), since it represents the minimum cost each agent would be willing to pay to transmit at the current time-slot. When the channel is occupied, no other transmission is allowed (line~\ref{alg:whittle-no-scheduling}).
The variable $ z $ keeps track of ongoing communication and drops to zero when a new transmission can be scheduled.
\begin{algorithm}
	\caption{Whittle Index Scheduling}
	\label{alg:optimal-scheduling}
	\begin{algorithmic}[1]
		\Require Processing time $\tau_i $, \commDel $ r_i(\cdot) $, and cost $ J_i(\cdot,\cdot) $
		for each agent $ i\in\sensSet $, time horizon $T$.
		\State $t = t_0$, $z = 0$;
		\While{$ t \le T $}
			\If{$z = 0$} \Comment{schedule transmission at time $ t $} 
				\State $ \displaystyle \pi \gets \argmax_{i\in\sensSet}\;W_i(A_i(t)) $; \Comment{trigger agent $ \pi $}\label{alg:whittle-scheduling}
				\State $ z \gets r_{\pi}(\tau_{\pi})-1 $;
			\Else	\Comment{continue ongoing transmission} \label{alg:whittle-no-scheduling}
				\State $ z \gets z-1 $;
			\EndIf
		\EndWhile
	\end{algorithmic}
\end{algorithm}

The Whittle index is known to be asymptotically optimal as $N \rightarrow \infty$, if a fluid limit condition is satisfied \cite{weber1990index, maatouk2020optimality}. These results, along with our simulations, suggest that the Whittle index is a very good low-complexity heuristic for scheduling in real-time monitoring and control applications.





\section{Applications}\label{sec:applications}

We demonstrate our co-design algorithms in 
two applications: multi-agent occupancy grid mapping in time-varying environments (Section~\ref{sec:sim-grid-mapping}), and 
ride sharing in autonomous vehicle networks (Section~\ref{sec:sim-ride-sharing}).
{The results show that we can achieve performance improvements of $18-35\%$ for grid mapping and $75-82\%$ for ride-sharing compared to baseline approaches.} We also provide a video briefly summarizing and visualizing our simulation results \cite{video}. 

\subsection{Multi-agent Mapping of Time-Varying Environments}
\label{sec:sim-grid-mapping}

\begin{figure}
\centering
\includegraphics[width=0.95\linewidth]{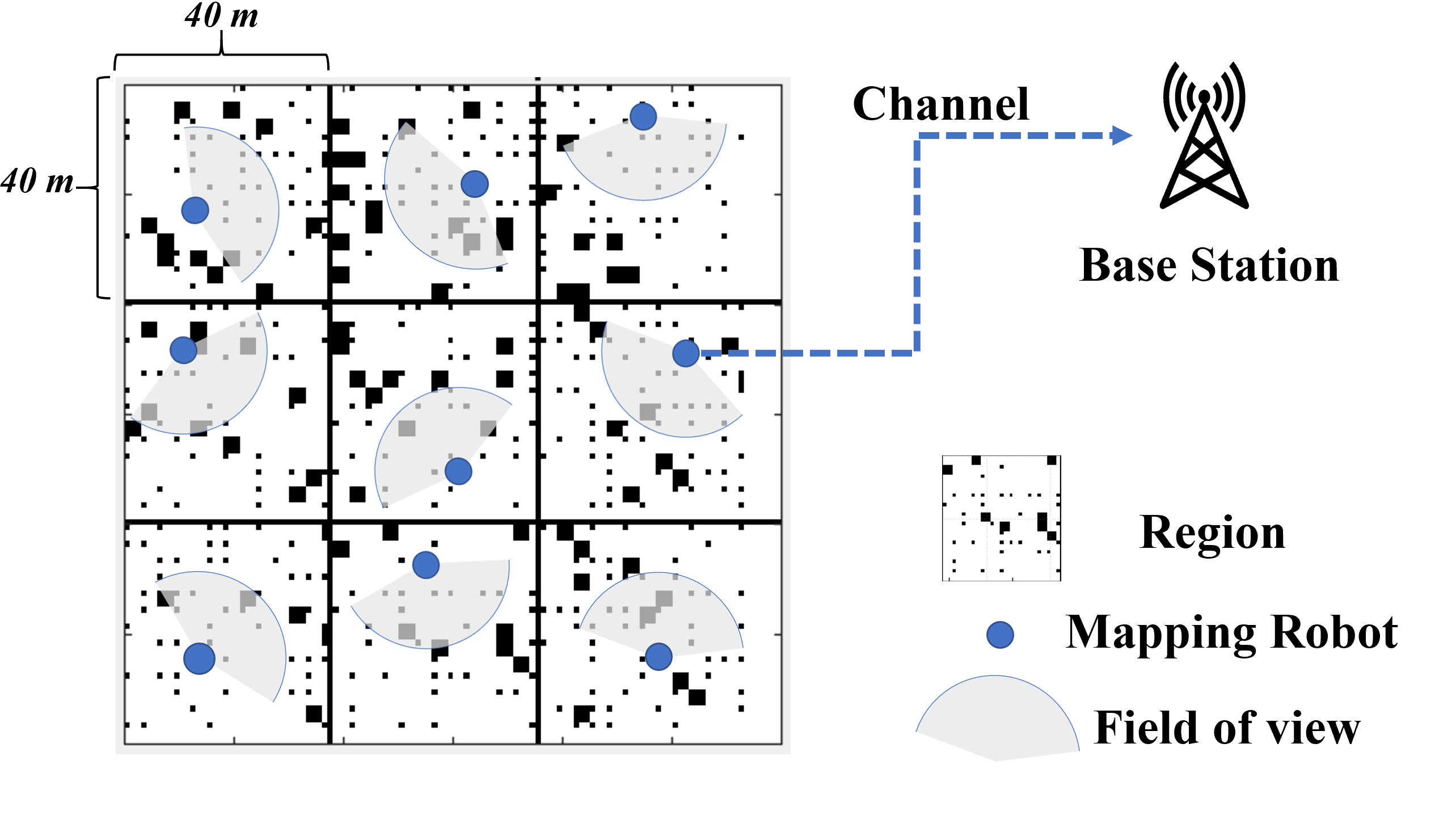}\vspace{-3mm}
\caption{{\bf Multi-agent mapping over 9 regions:} each agent monitors and builds a local grid map of a region, and sends map updates to a base station. The occupancy in the regions is time-varying.
A scheduling policy specifies how to share the communication channel among the agents. Processing times specify how much time each agent spends in generating new map updates.} 
\vspace{-3mm}
\label{figure:mapping_example}
\end{figure}

\myParagraph{Setup} We co-design computation and communication for a multi-agent mapping problem. We assume there are $N$ separate regions each of which is being mapped by an agent. The agents send updates ---in the form of occupancy grid maps of their surroundings--- to a base station over a single communication channel, where the local maps are aggregated into a global map for centralized monitoring (\autoref{figure:mapping_example}).

In our tests, each region is $40\mathrm{m}\times40\mathrm{m}$ in size and is represented by an occupancy grid map with $1\mathrm{m}\times1\mathrm{m}$ cells. The state of each cell can be either \textit{occupied} (1) or \textit{unoccupied} (0). 
We consider a dynamic environment where the state of each cell within region $i$ evolves according to a Markov chain, with cells remaining in their original state with probability $1-p_i$ and switching from occupied to unoccupied and vice-versa with probability $p_i$. This is a common model for grid mapping in dynamic environments in the robotics community \cite{meyer2012occupancy, saarinen2012independent}. 

Each agent is equipped with a range-bearing sensor (\eg lidar), with a fixed maximum scanning distance ($25\mathrm{m}$) and angular range $[-\pi/2, \pi/2]$. The agents move around the regions randomly, taking scans of the area round them. Scanning an entire region takes an agent multiple time-slots. We use the Navigation toolbox in MATLAB to create sensors such that the resolution of the readings $\theta_\text{min}$ improves with the processing time. We set $\theta_\text{min} = 0.5/\tau$. We also set the noise variance in angle and distance measurements to be inversely proportional to $\tau$. These settings capture the delay-accuracy trade-off. We further set the update communication times to increase linearly with the amount of processing, \ie $r(\tau) = 5 + \lceil \tau/2 \rceil$. 

The base station maintains an estimate of the current map for each region based on the most recent update it received and the Markov transition probabilities $\{p_i\}_{i \in \mathcal{V}}$ associated with each region.
As is common in mapping literature \cite{bourgault2002information,carrillo2015autonomous}, we measure uncertainty at the base station in terms of entropy of the current estimated occupancy grid map for each region and set the cost functions $J_i(\cdot,\cdot)$ to be the entropy of region $i$.
In~\cref{app:entropy}, we show that the entropy cost increases monotonically with the AoI of a region and satisfies the assumptions of our framework. It drops to a lower value if more time was spent in processing, since the base station is more certain about the quality of the received update. 
Our goal is to minimize the time-average of the entropies summed across each region through the joint optimization of processing times and the scheduling policy. 

\begin{figure}
\centering
\includegraphics[width=0.99\linewidth]{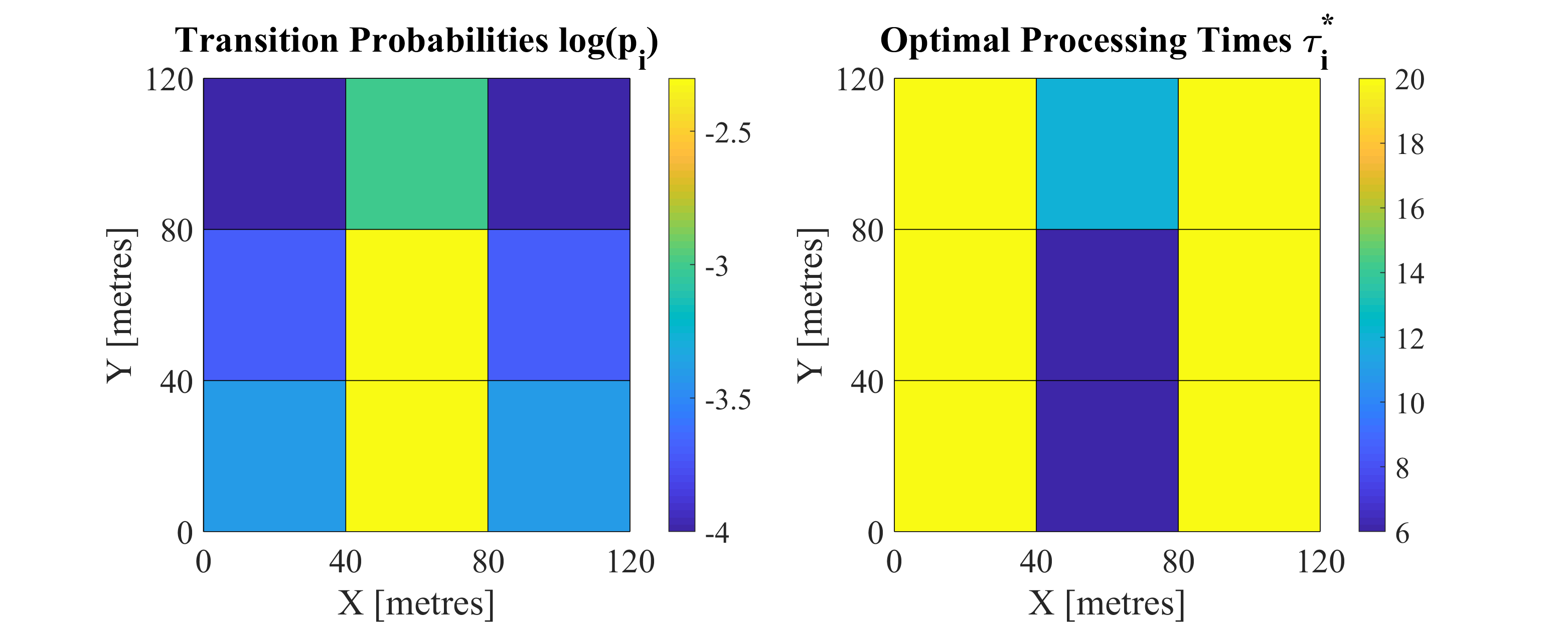}\vspace{-3mm}
\caption{Transition probabilities and optimal processing time allocations plotted for each region. The probabilities are plotted on a logarithmic scale while the processing times are plotted in number of time-slots.}
\vspace{-3mm}
\label{figure:opt_tau}
\end{figure}

\myParagraph{Results}
\autoref{figure:opt_tau} shows an example of transition probabilities $p_i$ (for each of the 9 regions) and the corresponding optimal processing times $\tau_i^*$ found using Algorithm \ref{alg:optimal-processing}. We observe that for regions that change quickly (\ie have large value of $p_i$), the corresponding processing time allocated is smaller. This is because there is not much benefit to spending large amounts of time generating high quality updates if they become outdated very quickly. Conversely, for slowly changing regions (with low values of $p_i$), Algorithm \ref{alg:optimal-processing} assigns much longer processing times. In this case, high quality useful updates can be created by taking longer time since the regions don't change quickly.

\begin{figure}
\centering
\includegraphics[width=1.0\linewidth]{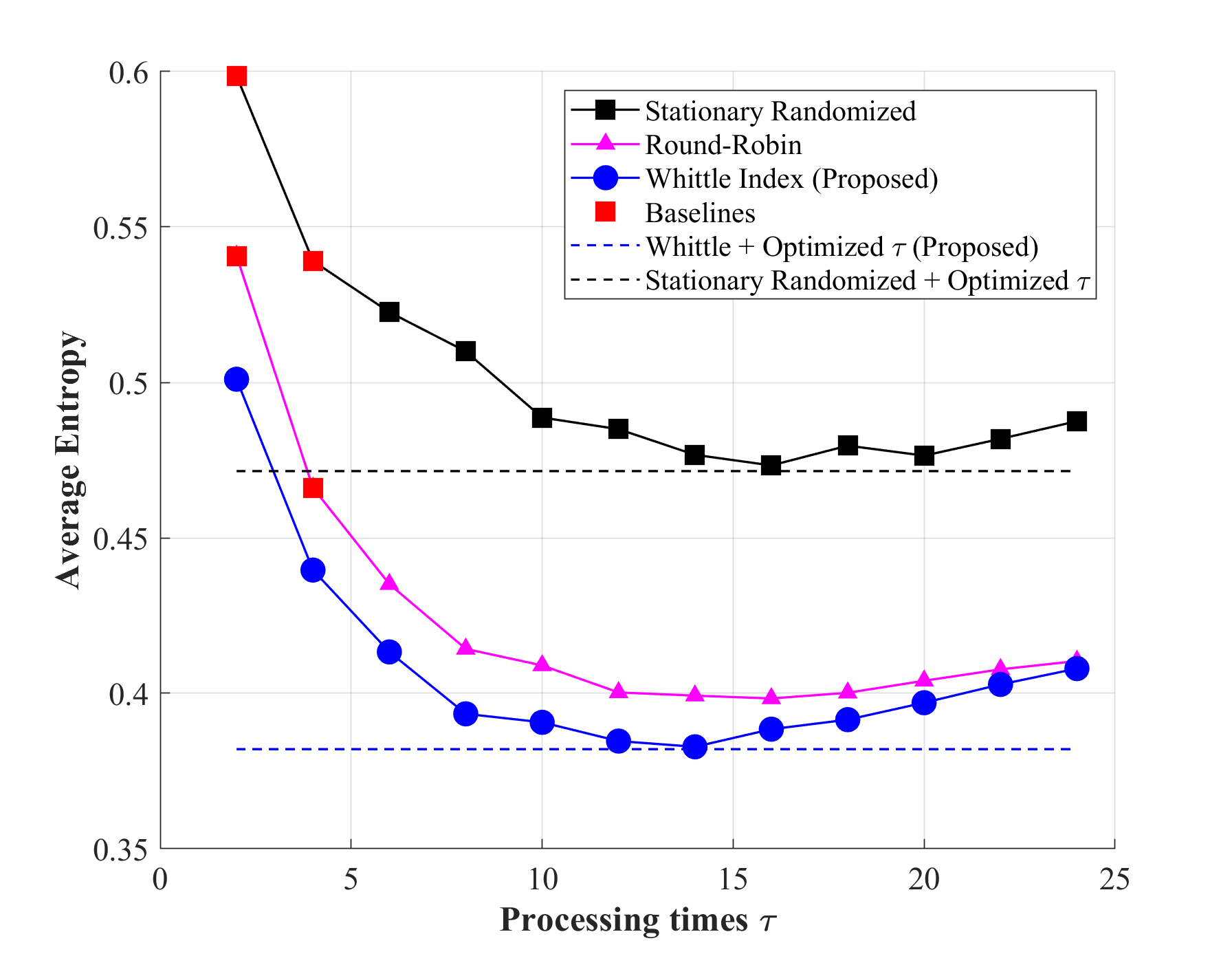}\vspace{-3mm}
\caption{Performance of different scheduling policies vs. processing times $\tau$. Solid lines represent performance of different classes of scheduling policies as the processing time $\tau$ varies. The dotted lines represent the scheduling performance with processing times computed using Algorithm \ref{alg:optimal-processing}.}
\vspace{-3mm}
\label{figure:mapping_results}
\end{figure}

Further, we compare the performance of various scheduling algorithms in~\autoref{figure:mapping_results}.
We consider the setting where the processing times $\tau_i$ are fixed to be the same parameter $\tau$ for every region (uniform processing allocation).
We then plot the performance of three scheduling algorithms --a uniform stationary randomized policy, a round-robin policy, and the proposed Whittle index-based policy-- for different values of $\tau$. We also plot the performance of the Whittle index policy and the stationary randomized policy under the optimized processing times, computed using Algorithm \ref{alg:optimal-processing}, shown via dotted lines in \autoref{figure:mapping_results}. We observe that Algorithm~\ref{alg:optimal-processing} can find processing times that perform well in practice. We also observe that the Whittle index policy outperforms the two ``traditional'' classes of scheduling policies for every value of the parameter $\tau$. 


Overall, choosing the processing times using Algorithm \ref{alg:optimal-processing} and using the Whittle schedule from Algorithm \ref{alg:optimal-scheduling} together leads to a performance improvement of $28-35\%$ over the baseline versions of randomized policies. Similarly, our proposed approach leads to a performance improvement of $17-28\%$ over the baseline versions of round-robin policies.



\subsection{Smart Ride Sharing Control in Vehicle Networks}
\label{sec:sim-ride-sharing}

\myParagraph{setup}
We consider the scenario in which a ride-sharing taxi fleet serves a city
coordinated by a central scheduler,
which receives riding requests and assigns them
to the drivers. 
Assigned requests are enqueued
into a FIFO-like queue for each driver.
In particular,
a rider is matched to the driver whose predicted route has
the shortest distance to the pick-up location.

In our setup, routes are calculated locally by drivers and transmitted on demand to the scheduler,
which uses this information to match future requests.
Such distributed processing for route optimization is different from current architectures, which are usually centralized.
However, it allows for much greater scalability and is envisioned as a key component in increasing efficiency and scale of future ride-sharing systems~\cite{7393588,deeppool,muelas2015distributed}.

Given communication constraints,
only one driver can transmit at a time. 
Drivers update their route periodically
to embed real-time road conditions
and remove served requests from the queue.
Routes are calculated via the Travelling Salesman Problem (TSP)
involving the first $ R $ pick-up and drop-off locations in the request queue (\autoref{fig:request-queue-optimization}).
Processing many requests ensures more efficient paths for enqueued riders,
thus shortening their \emph{travel time} from pick up to drop off.
Conversely, the complexity of the TSP (\ie its \compDel)
increases with the amount of processed requests $ R $.
As a consequence, the information collected by the scheduler
is usually older, 
inducing larger gaps with the actual route followed by the driver (\autoref{fig:request-matching}).
This leads to worse driver-request matching
and increases the \emph{waiting time} experienced by riders before they are actually picked up.
Since the overall Quality of Service (QoS) is measured through the \emph{service time},
given by the sum of travel and waiting times of riders,
the drivers face a trade-off:
processing many requests shortens the travels,
while processing few reduces the waiting time.

\begin{figure}
	\centering
	\begin{tikzpicture}
	\footnotesize
	\tikzset{
		state/.style={
			rectangle split,
			rectangle split,
			rectangle split parts=2,
			draw=black,
			ultra thin,
			minimum width=.5cm,
			inner sep=2pt,
			text centered,
			append after command={(\tikzlastnode.south east) edge[to path={rectangle (\tikztotarget)}, thick] (\tikzlastnode.north west)}
		}
	}
	\node[state,rectangle split part fill={green!30,green!50}] at (0,0) (a) 
			{$ 1_P $ \nodepart{two} $ 1_D $};
	\node[state,rectangle split part fill={green!30,green!50}] (b) [left of=a,node distance=.5cm] 
			{$ 2_P $ \nodepart{two} $ 2_D $};
	\node[state,rectangle split part fill={black!10,black!20}] (c) [left of=b,node distance=.5cm] 
			{$ 3_P $ \nodepart{two} $ 3_D $};
	\node[state,rectangle split part fill={black!10,black!20}] (d) [left of=c,node distance=.5cm] 
		{$ 4_P $ \nodepart{two} $ 4_D $};
	\node[state,rectangle split part fill={black!10,black!20}] (e) [left of=d,node distance=.5cm] 
		{$ 5_P $ \nodepart{two} $ 5_D $};
	\draw[ultra thick] (-2.8,-.38) -- (.25,-.38) -- (.25,.38) -- (-2.8,.38);
	\node at (-1.25,.7) {Request queue};
	\node[state,rectangle split part fill={black!10,black!20}] (f) [left of=e,node distance=2cm] 
		{$ 6_P $ \nodepart{two} $ 6_D $};
	\draw[->] (-3.4,0) -- (-3,0);
	
	\draw[->] (-.25,-.5) -- (-.25,-.8);
	\node[draw,thick,rectangle,fill=green!40] at (-.25,-1.2) (TSP) {\scriptsize\shortstack{TSP \\ $ R\!=\!2 $}};
	\draw[->] (.3,-1.2) -- (.8,-1.2);
	
	\node[draw,rectangle,minimum height=3cm,minimum width=3cm] at (2.5cm,0) (map) {};
	\foreach \x in {1,1.3,...,4}
	{	\draw[ultra thin] (\x,-1.5) -- (\x,1.5);
	}
	\foreach \x in {-1.5,-1.2,...,1.5}
	{	\draw[ultra thin] (1,\x) -- (4,\x);
	}
	
	\draw[fill] (3.5,-.5cm) circle [radius=0.1cm];
	\node at (3.5,-.5cm) (start) {};
	
	\draw[fill=green!30] (2.75,-0.75cm) circle [radius=0.2cm];
	\node at (2.75,-0.75cm) (1p) {\scriptsize$ 1_P $};
	
	\draw[fill=green!50] (2.4,1.2cm) circle [radius=0.2cm];
	\node at (2.4,1.2cm) (1d) {\scriptsize$ 1_D $};
	
	\draw[fill=green!30] (1.3,-0.5cm) circle [radius=0.2cm];
	\node at (1.3,-0.5cm) (2p) {\scriptsize$ 2_P $};
	
	\draw[fill=green!50] (1.7,1.1cm) circle [radius=0.2cm];
	\node at (1.7,1.1cm) (2d) {\scriptsize$ 2_D $};
	
	\path[->,shorten >= -2pt, shorten <= -2pt,thick] (start) edge (1p) (1p) edge (2p) (2d) edge (1d);
	\draw[->,thick] (2p) edge (2d);
	
	\draw[fill=black!10] (3,.2cm) circle [radius=0.2cm];
	\node at (3,.2cm) (3p) {\scriptsize$ 3_P $};
	\draw[fill=black!30] (3.7,1.1cm) circle [radius=0.2cm];
	\node at (3.7,1.1cm) (3d) {\scriptsize$ 3_D $};
	\draw[fill=black!10] (1.25,-1cm) circle [radius=0.2cm];
	\node at (1.25,-1cm) (3p) {\scriptsize$ 4_P $};
	\draw[fill=black!30] (3.6,.2cm) circle [radius=0.2cm];
	\node at (3.6,.2cm) (4d) {\scriptsize$ 4_D $};
	\draw[fill=black!10] (3.8,-1.2cm) circle [radius=0.2cm];
	\node at (3.8,-1.2cm) (5p) {\scriptsize$ 5_P $};
	\draw[fill=black!30] (1.25,0cm) circle [radius=0.2cm];
	\node at (1.25,0cm) (5p) {\scriptsize$ 5_D $};
\end{tikzpicture}
	\caption{Drivers calculate their route by processing the oldest $ R $ requests (green queue portion).
		The TSP solver starts
		from the current driver location 
		and involves pick ups (P) and drop offs (D) of the processed requests.
		\label{fig:request-queue-optimization}
	}
\end{figure}
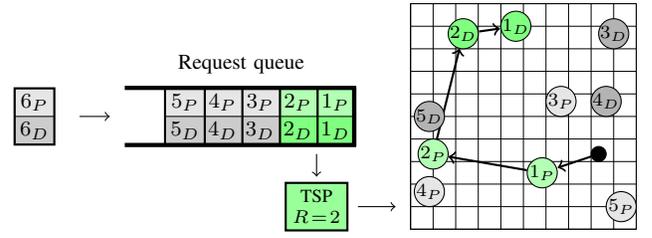
\begin{figure}
	\centering
	\begin{minipage}[l]{.5\linewidth}
		\centering
		\scalebox{1}{
    \begin{tikzpicture}
    	\tikzset{
    		stop/.style={
    			fill,
    			circle,
    			scale=.5}
    	}
    	\node[draw,rectangle,minimum height=3cm,minimum width=3cm] at (0,0) (map) {};
    	\foreach \x in {-1.5,-1.2,...,1.5}
    	{	\draw[ultra thin] (\x,-1.5) -- (\x,1.5);
    	}
    	\foreach \x in {-1.5,-1.2,...,1.5}
    	{	\draw[ultra thin] (-1.5,\x) -- (1.5,\x);
    	}
    	
    	\node[stop] at (1.2,-1.4) (0) {};
    	\node[stop] at (1,-0.75) (1) {};
    	\node[stop] at (.5,-0.8) (2) {};
    	\path[->,thick] (0) edge (1) (1) edge (2);
    	
    	\node[stop,fill=black!50] at (.2,-.1) (3) {};
    	\node[stop,fill=black!50] at (1,.6) (4) {};
    	\node[stop,fill=black!50] at (1.3,1.2) (5) {};
    	\node[stop,fill=black!50] at (.5,1.3) (6) {};
    	\path[->,color=black!50,thick] (2) edge (3) (3) edge (4) (4) edge (5) (5) edge (6);
    	
    	\node[stop,fill=black!50] at (-1,0) (41) {};
    	\node[stop,fill=black!50] at (-1.3,-1) (51) {};
    	\path[->,dashed,color=black!50,thick] (3) edge (41) (41) edge (51);
    	
    	\node[stop,fill=red] at (-.8,-1.35)(req) {};
    	\node[stop,fill=red] at (-1.3,.4)(req) {};
    \end{tikzpicture}
}
		\label{fig:request-matching-bad}
	\end{minipage}%
	\begin{minipage}[r]{.5\linewidth}
		\centering
		\scalebox{1}{
    \begin{tikzpicture}
    	\tikzset{
    		stop/.style={
    			fill,
    			circle,
    			scale=.5}
    	}
    	\node[draw,rectangle,minimum height=3cm,minimum width=3cm] at (0,0) (map) {};
    	\foreach \x in {-1.5,-1.2,...,1.5}
    	{	\draw[ultra thin] (\x,-1.5) -- (\x,1.5);
    	}
    	\foreach \x in {-1.5,-1.2,...,1.5}
    	{	\draw[ultra thin] (-1.5,\x) -- (1.5,\x);
    	}
    	
    	\node[stop] at (1.2,-1.4) (0) {};
    	\node[stop] at (1,-0.75) (1) {};
    	\node[stop] at (.5,-0.8) (2) {};
    	\node[stop] at (.2,-.1) (3) {};
    	\node[stop] at (1,.6) (4) {};
    	\path[->,thick] (0) edge (1) (1) edge (2) (2) edge (3) (3) edge (4);
    	
    	\node[stop,fill=black!50] at (1.3,1.2) (5) {};
    	\node[stop,fill=black!50] at (.5,1.3) (6) {};
    	\path[->,color=black!50,thick] (4) edge (5) (5) edge (6);
    	
    	\node[stop,fill=black!50] at (.1,.9) (51) {};
    	\path[->,dashed,color=black!50,thick] (4) edge (51);
    	
    	\node[stop,fill=green] at (-.5,1) (req) {};
    	\node[stop,fill=green] at (.2,.4)(req) {};
    \end{tikzpicture}
}
		\label{fig:request-matching-good}
	\end{minipage}
	\caption{Left: long processing may cause large gaps between 
	    new routes calculated by the driver (solid gray) 
	    and the outdated ones stored at the scheduler (dashed gray), yielding bad matches (red dots).
		Right: with short processing, the matched requests are close to the actual routes (green dots).
		\label{fig:request-matching}
		\vspace{-5mm}
	}
\end{figure}
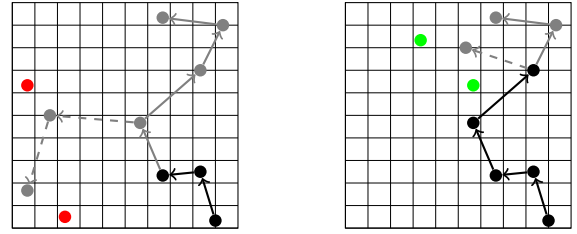

In our tests, we model the city as a 200-node graph where each driver travels
one edge per time slot.
Requests are randomly generated according to a Poisson process of unit intensity
and assigned immediately to the matching driver by the scheduler.
Each request contributes one time slot to the processing time of the TSP
(\eg $\delayComp=2$ corresponds to processing two requests)
and we set $ \delayComm = \delayComp $
(longer processing yields longer routes to transmit).
To exploit the advantage of the Whittle index,
we simulate an heterogeneous fleet with
five ``myopic'' drivers, which can only process the oldest request ($ \delayComp_m = 1 $),
and five ``smart'' drivers 
whose processing can be designed:
in particular, we assign the same \compDel $ \delayComp_s $ 
to all such ``smart" drivers.
The cost of each driver, given by its \emph{average service time} (AST), is modeled as
\begin{equation}\label{eq:ride-sharing-cost}
	J_i\left(\delayComp_i,A_i(t)\right) = P_i(\delayComp_i) + A_i(t)
\end{equation}
where the estimated contribution of the local processing (TSPs)
\begin{equation}\label{eq:ride-sharing-processing-cost}
	P_i(\delayComp_i) \doteq \left(2 + 2\mbox{e}^{-0.2\delayComp_i}\right)\hat{q}_i(t)
\end{equation}
was fitted from simulations with an initial queue and no assignments. 
Because the number of enqueued requests affects the AST but cannot be computed offline,
we modeled $ P_i(\delayComp_i) $ as linear with the queue length.
The scheduler approximates the queue length at time $ t $
with the latest received value $ \hat{q}_i(t) $.
The dependence on $ A_i(t) $ is hard to assess and we let it linear.
\footnote{Other cost functions decreasing with $ \delayComp_i $ and increasing with $ A_i(t) $ also yield good performance,
	suggesting that our approach is indeed robust.
}

\begin{figure}
	\centering
	\begin{tikzpicture}[every pin/.style={draw,fill=white,pin edge={black},pin distance=.5cm,minimum width=3}]
	\centering
	
	\begin{axis}[grid,scale=.8,
				xlabel={Processing time of smart drivers $ \delayComp_s $},
				xtick={1,2,...,7},
				xlabel style={yshift=2pt},
				ylabel={Average service time},
				ytick={30,70,...,450},
				tick label style={font=\small},
				label style={font=\small},
				legend style={at={(0.01,.99)},anchor=north west,legend cell align={left},font=\scriptsize},
				width=1.2\linewidth,
				height=.8\linewidth]
		\addplot[color=black,mark=square*,mark size=10/4] table [col sep=comma] {Images/plot_data/rand_1000.csv};
		\addplot[color=blue,mark=*,mark size=10/4] table [col sep=comma] {Images/plot_data/whittle_lin_1000.csv};
		\addplot[color=red,mark=square*,mark size=10/4,only marks,restrict expr to domain={\coordindex}{0:1}] table [col sep=comma] {Images/plot_data/rand_1000.csv};
		\addplot[color=green,mark=*,mark size=10/4,only marks,restrict expr to domain={\coordindex}{4:4}] table [col sep=comma] {Images/plot_data/whittle_lin_1000.csv};
		\legend {Stationary Randomized, Whittle index (proposed), Baselines, Whittle index + optimized processing (proposed)};
	
	\coordinate (pt) at (axis cs:5,48);
	
	\end{axis}

	\node[rectangle,draw,minimum size=10,pin=93:{%
		\begin{tikzpicture}[trim axis right,scale=.2]
			\begin{axis}[
				xmin=5,
				xmax=5,
				xtick={4,5,6},
				xticklabel style={yshift=-7pt},
				ymin=40,
				ytick={40,45,50,55},
				yticklabel style={xshift=-5pt},
				tick label style={font={\Huge \boldmath}},
				grid,
				enlargelimits,
				]
				\addplot[color=black,mark=square*,mark size=10,restrict expr to domain={\coordindex}{4:4}] table [col sep=comma] {Images/plot_data/rand_1000.csv};
				\addplot[draw=blue,fill=green,mark=*,mark size=12,restrict expr to domain={\coordindex}{4:4}] table [col sep=comma] {Images/plot_data/whittle_lin_1000.csv};
			\end{axis}
		\end{tikzpicture}%
	}] at (pt) {} ;

\end{tikzpicture}\vspace{-4mm}
	\caption{Average service time with varying \compDel $ \delayComp_s $.}
	\vspace{-4mm}
	\label{fig:ride-sharing-performance}
\end{figure}
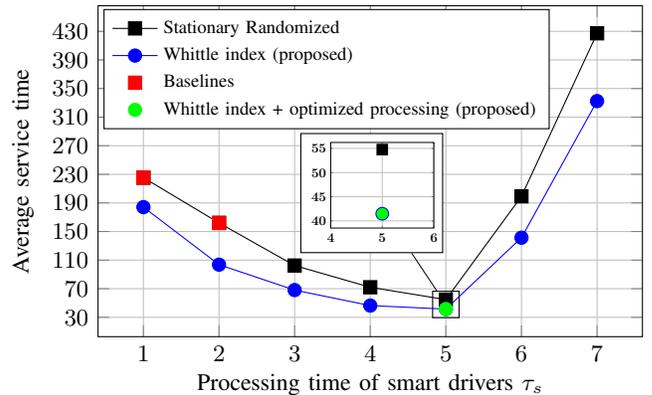
\myParagraph{results}
We compute statistics over $ 1000 $ Monte Carlo runs.
~\autoref{fig:ride-sharing-performance} shows the AST
with $ 10000 $ requests assigned during the simulation
for $ \delayComp_s\in\{1,...,7\} $.
The circles refer to the performance obtained with the Whittle index policy,
while the squares to Stationary Randomized which is used as a benchmark.
Combining Whittle index-based scheduling
with processing optimization (green circle)
yields a striking improvement of the QoS (AST $=41 $)
compared to the Stationary Randomized with standards policies (red squares) such as
FIFO request service ($ \delayComp_s = 1 $, AST $=225$),
or back-to-back trips~\cite{uberSharedRides} ($ \delayComp_s = 2 $, AST $=165$).
In particular, the minimum at $ \delayComp_s^*=5 $
indicates that it is optimal to process the five oldest requests in the queue. 
Also, the Whittle index outperforms Stationary Randomized for all values of the \compDel,
with a decrease at the optimum of $ 25\% $.	

\section{Conclusion}\label{sec:conclusion}


In this work, we developed a novel framework for computation and communication co-design for real-time multi-agent monitoring and control.
We designed efficient algorithms that jointly allocate the processing time for each agent and schedule the available network communication resources.
Through simulations, we further demonstrated that the proposed approach works well for two different applications: multi-agent occupancy grid mapping in time-varying environments and distributed ride sharing in autonomous vehicle networks.

Possible directions of future work involve extending the theoretical framework to consider more complex and realistic cost functions that are
coupled across multiple agents, time-varying or unknown, requiring learning-based approaches. 

\bibliographystyle{IEEEtran}
\bibliography{bibfile}

\begin{thebibliography}{10}
\providecommand{\url}[1]{#1}
\csname url@samestyle\endcsname
\providecommand{\newblock}{\relax}
\providecommand{\bibinfo}[2]{#2}
\providecommand{\BIBentrySTDinterwordspacing}{\spaceskip=0pt\relax}
\providecommand{\BIBentryALTinterwordstretchfactor}{4}
\providecommand{\BIBentryALTinterwordspacing}{\spaceskip=\fontdimen2\font plus
\BIBentryALTinterwordstretchfactor\fontdimen3\font minus
  \fontdimen4\font\relax}
\providecommand{\BIBforeignlanguage}[2]{{%
\expandafter\ifx\csname l@#1\endcsname\relax
\typeout{** WARNING: IEEEtran.bst: No hyphenation pattern has been}%
\typeout{** loaded for the language `#1'. Using the pattern for}%
\typeout{** the default language instead.}%
\else
\language=\csname l@#1\endcsname
\fi
#2}}
\providecommand{\BIBdecl}{\relax}
\BIBdecl

\bibitem{kaul2012real}
S.~Kaul, R.~Yates, and M.~Gruteser, ``Real-time status: How often should one
  update?'' in \emph{Proc. IEEE INFOCOM}, 2012, pp. 2731--2735.

\bibitem{yin17_tit_update_or_wait}
Y.~Sun, E.~Uysal-Biyikoglu, R.~D. Yates, C.~E. Koksal, and N.~B. Shroff,
  ``Update or wait: How to keep your data fresh,'' \emph{IEEE Trans.
  Information Theory}, vol.~63, no.~11, pp. 7492--7508, Nov. 2017.

\bibitem{kadota2018scheduling}
I.~Kadota, A.~Sinha, E.~Uysal-Biyikoglu, R.~Singh, and E.~Modiano, ``Scheduling
  policies for minimizing age of information in broadcast wireless networks,''
  \emph{IEEE/ACM Trans. Netw.}, vol.~26, no.~6, pp. 2637--2650, 2018.

\bibitem{talak2018optimizing}
R.~Talak, S.~Karaman, and E.~Modiano, ``Optimizing information freshness in
  wireless networks under general interference constraints,'' in \emph{Proc.
  ACM Int. Symp. Mobile Ad Hoc Netw. Comput. (MobiHoc)}, 2018, pp. 61--70.

\bibitem{tripathi2019whittle}
V.~Tripathi and E.~Modiano, ``A whittle index approach to minimizing functions
  of age of information,'' in \emph{Proc. 57th Allerton Conf. Commun. Control
  Comput.}\hskip 1em plus 0.5em minus 0.4em\relax IEEE, 2019, pp. 1160--1167.

\bibitem{xiao2005scheme}
L.~Xiao, S.~Boyd, and S.~Lall, ``A scheme for robust distributed sensor fusion
  based on average consensus,'' in \emph{IPSN 2005. Fourth Int. Symp. Inf.
  Proc. Sensor Netw., 2005}, 2005, pp. 63--70.

\bibitem{carli2008distributed}
R.~Carli, A.~Chiuso, L.~Schenato, and S.~Zampieri, ``Distributed kalman
  filtering based on consensus strategies,'' \emph{IEEE Journal on Selected
  Areas in communications}, vol.~26, no.~4, pp. 622--633, 2008.

\bibitem{olfati2005consensus}
R.~Olfati-Saber and J.~S. Shamma, ``Consensus filters for sensor networks and
  distributed sensor fusion,'' in \emph{Proc. 44th IEEE CDC}, 2005, pp.
  6698--6703.

\bibitem{sun2017remote}
Y.~Sun, Y.~Polyanskiy, and E.~Uysal-Biyikoglu, ``Remote estimation of the
  wiener process over a channel with random delay,'' in \emph{Proc. IEEE Int.
  Symp. Information Theory (ISIT)}, 2017, pp. 321--325.

\bibitem{ornee2019sampling}
T.~Z. Ornee and Y.~Sun, ``Sampling for remote estimation through queues: Age of
  information and beyond,'' \emph{IEEE Int. Symp. Model. Optim. Mobile, Ad Hoc
  Wireless Netw. (WiOpt)}, 2019.

\bibitem{champati2019performance}
J.~P. Champati, M.~H. Mamduhi, K.~H. Johansson, and J.~Gross, ``Performance
  characterization using aoi in a single-loop networked control system,'' in
  \emph{Proc. IEEE INFOCOM AoI Workshop}, 2019, pp. 197--203.

\bibitem{klugel2019aoi}
M.~Kl{\"u}gel, M.~H. Mamduhi, S.~Hirche, and W.~Kellerer, ``Aoi-penalty
  minimization for networked control systems with packet loss,'' in \emph{Proc.
  IEEE INFOCOM AoI Workshop}, 2019, pp. 189--196.

\bibitem{kosta2017age}
A.~Kosta, N.~Pappas, V.~Angelakis \emph{et~al.}, ``Age of information: A new
  concept, metric, and tool,'' \emph{Foundations and Trends in Networking},
  vol.~12, no.~3, pp. 162--259, 2017.

\bibitem{sun2019age_book}
Y.~Sun, I.~Kadota, R.~Talak, and E.~Modiano, ``Age of information: A new metric
  for information freshness,'' \emph{Synthesis Lectures on Communication
  Networks}, vol.~12, no.~2, pp. 1--224, 2019.

\bibitem{NNRuntime}
M.~{Amir} and T.~{Givargis}, ``Priority neuron: A resource-aware neural network
  for cyber-physical systems,'' \emph{IEEE Trans. Comp.-Aided Design of
  Integrated Circ. and Sys.}, vol.~37, no.~11, pp. 2732--2742, Nov 2018.

\bibitem{sandler2018mobilenetv2}
M.~Sandler, A.~Howard, M.~Zhu, A.~Zhmoginov, and L.-C. Chen, ``Mobilenetv2:
  Inverted residuals and linear bottlenecks,'' in \emph{Proc. IEEE CVPR}, 2018,
  pp. 4510--4520.

\bibitem{2018arXiv180402767R}
J.~{Redmon} and A.~{Farhadi}, ``{YOLOv3: An Incremental Improvement},''
  \emph{arXiv e-prints}, p. arXiv:1804.02767, Apr 2018.

\bibitem{howard2019searching}
A.~Howard, M.~Sandler, G.~Chu, L.-C. Chen, B.~Chen, M.~Tan, W.~Wang, Y.~Zhu,
  R.~Pang, V.~Vasudevan \emph{et~al.}, ``Searching for mobilenetv3,'' in
  \emph{Proc. IEEE/CVF Int. Conf. Comp. Vision}, 2019, pp. 1314--1324.

\bibitem{crankshaw2017clipper}
D.~Crankshaw, X.~Wang, G.~Zhou, M.~J. Franklin, J.~E. Gonzalez, and I.~Stoica,
  ``Clipper: A low-latency online prediction serving system,'' in \emph{14th
  USENIX Symposium on Networked Systems Design and Implementation (NSDI 17)},
  2017, pp. 613--627.

\bibitem{Pavone-RSS-19}
S.~Chinchali, A.~Sharma, J.~Harrison, A.~Elhafsi, D.~Kang, E.~Pergament,
  E.~Cidon, S.~Katti, and M.~Pavone, ``Network offloading policies for cloud
  robotics: A learning-based approach,'' in \emph{Proceedings of Robotics:
  Science and Systems}, FreiburgimBreisgau, Germany, June 2019.

\bibitem{ballotta2019computationcommunication}
L.~{Ballotta}, L.~{Schenato}, and L.~{Carlone}, ``Computation-communication
  trade-offs and sensor selection in real-time estimation for processing
  networks,'' \emph{IEEE Trans. Net. Sci. Eng.}, vol.~7, no.~4, 2020.

\bibitem{whittle1988restless}
P.~Whittle, ``Restless bandits: Activity allocation in a changing world,''
  \emph{Journal of applied probability}, pp. 287--298, 1988.

\bibitem{weber1990index}
R.~R. Weber and G.~Weiss, ``On an index policy for restless bandits,''
  \emph{Journal of applied probability}, pp. 637--648, 1990.

\bibitem{maatouk2020optimality}
A.~Maatouk, S.~Kriouile, M.~Assaad, and A.~Ephremides, ``On the optimality of
  the whittle’s index policy for minimizing the age of information,''
  \emph{IEEE Trans. Wireless Commun.}, 2020.

\bibitem{video}
\BIBentryALTinterwordspacing
V.~Tripathi, L.~Ballotta, L.~Carlone, and E.~Modiano, ``Computation and
  communication co-design for real-time monitoring and control in multi-agent
  systems,'' Video Attachment, 2021. [Online]. Available:
  \url{https://www.dropbox.com/s/q7ijfsfc6eoko9d/video_hq.mp4}
\BIBentrySTDinterwordspacing

\bibitem{meyer2012occupancy}
D.~Meyer-Delius, M.~Beinhofer, and W.~Burgard, ``Occupancy grid models for
  robot mapping in changing environments,'' in \emph{Proceedings of the AAAI
  Conference on Artificial Intelligence}, vol.~26, no.~1, 2012.

\bibitem{saarinen2012independent}
J.~Saarinen, H.~Andreasson, and A.~J. Lilienthal, ``Independent markov chain
  occupancy grid maps for representation of dynamic environment,'' in
  \emph{IEEE/RSJ Int. Conf. Intell. Robot. Syst.}, 2012, pp. 3489--3495.

\bibitem{bourgault2002information}
F.~Bourgault, A.~A. Makarenko, S.~B. Williams, B.~Grocholsky, and H.~F.
  Durrant-Whyte, ``Information based adaptive robotic exploration,'' in
  \emph{IEEE/RSJ Int. Conf. Intell. Robot. Syst.}, vol.~1, 2002, pp. 540--545.

\bibitem{carrillo2015autonomous}
H.~Carrillo, P.~Dames, V.~Kumar, and J.~A. Castellanos, ``Autonomous robotic
  exploration using occupancy grid maps and graph slam based on shannon and
  r{\'e}nyi entropy,'' in \emph{IEEE ICRA}, 2015, pp. 487--494.

\bibitem{7393588}
A.~Y.~S. {Lam}, Y.~{Leung}, and X.~{Chu}, ``Autonomous-vehicle public
  transportation system: Scheduling and admission control,'' \emph{IEEE Trans.
  Intell. Transp. Syst.}, vol.~17, no.~5, pp. 1210--1226, 2016.

\bibitem{deeppool}
A.~O. {Al-Abbasi}, A.~{Ghosh}, and V.~{Aggarwal}, ``Deeppool: Distributed
  model-free algorithm for ride-sharing using deep reinforcement learning,''
  \emph{IEEE Trans. Intell. Transp. Syst.}, vol.~20, no.~12, pp. 4714--4727,
  2019.

\bibitem{muelas2015distributed}
S.~Muelas, A.~LaTorre, and J.-M. Pena, ``A distributed vns algorithm for
  optimizing dial-a-ride problems in large-scale scenarios,''
  \emph{Transportation Research Part C: Emerging Technologies}, vol.~54, pp.
  110--130, 2015.

\bibitem{uberSharedRides}
\BIBentryALTinterwordspacing
Uber. (2018) How does uber pool expand access? [Online]. Available:
  \url{www.uber.com/us/en/marketplace/matching/shared-rides/}
\BIBentrySTDinterwordspacing

\end{thebibliography}

\appendix

\subsection{Proof of~\cref{thm:optimal-threshold}}
\label{app:threshold-policy}
We first establish that the decoupled \cref{prob:decoupled-problem} is equivalent to a Markov decision process (MDP). We then solve the MDP using dynamic programming. Since the analysis looks similar for each of the $N$ decoupled problems, we drop the subscript $i$ and solve the problem for a generic agent.

The state of the MDP describing \cref{prob:decoupled-problem} consists of two non-negative integers $\big(A(t),z(t) \big)$. $A(t)$ denotes the AoI of the agent at time $t$ while $z(t)$ denotes how much time is left in the ongoing transmission from this agent. When the agent is not transmitting, $z(t)$ is set to be $0$.

The variable $u(t)$ is an indicator variable that denotes the action of the agent: whether it is transmitting in time-slot $t$ or not. Its value is chosen from the action set $\{0,1\}$, $0$ meaning the agent is at rest and $1$ meaning an ongoing transmission. When $z(t) > 0$, that means a transmission is ongoing and $u(t)$ can only be set to $1$. This ensures that an entire update must be finished by the agent before making the next scheduling decision. Whenever $z(t) = 0$, the scheduler can choose $u(t)$ to be either $0$ or $1$, indicating the beginning of a new transmission. 

The MDP evolution can be split into 2 cases. When the agent is not transmitting  ($u(t) = 0$), AoI increases by $1$ and $z(t)$ remains at $0$. 
\begin{equation}\label{eq:state-dynamics-1}
	\big(A(t+1),z(t+1)\big)_{u=0} = (A(t)+1,0).
\end{equation}
When the agent is transmitting ($u(t) = 1$), the AoI drops when a new update completes delivery. Otherwise, it keeps increasing by $1$. The variable $z(t)$ is set to $r(\tau)-1$ at the beginning of a new transmission to indicate the time left in completing it. It decreases by $1$ in every time-slot thereon, until the transmission completes and $z$ becomes $0$. Thus, the state evolution is given by:
\begin{equation}
\big(z(t+1)\big)_{u=1} =
 \begin{cases}
      r(\tau) - 1, &  \text{ if } z(t) = 0 \\
      z(t)-1, & \text{ otherwise.}
 \end{cases}
\end{equation}
\begin{equation}
\big(A(t+1)\big)_{u=1} =
 \begin{cases}
      \Delta, &  \text{ if } z(t+1) = 0.\\
      A(t)+1, & \text{ otherwise. }
 \end{cases}
\end{equation}

Now that we have specified the state space, the action space and the evolution equations; we also need to specify a cost function. We assume that in each time-slot the scheduler pays a cost of the form $Cu(t) + J(\tau, A(t))$. This maps the current state and action to a cost, where $C$ acts like a transmission charge and $J(\tau, A(t))$ is an increasing function of the AoI, given a fixed value of $\tau$. 

Note that the decision process we have set up above is Markov since the state evolution depends only on the states and the actions taken in the previous time-slot. We wouldn't have been able to make this conclusion without assuming a fixed value of the waiting times $\delta_i$, since that would have required us to maintain history per update.

Next, we aim to minimize the infinite horizon time-average cost for this MDP using dynamic programming. We follow the standard approach by first setting up the Bellman recursions. The case when $r(\tau) = 1$ is a direct application of Theorem 1 in \cite{tripathi2019whittle}, but with an adjusted minimum AoI value. For the discussion that follows, we assume the more interesting case of $r(\tau) > 1$.

We start from a state where the AoI $A(t) = h$ and there is no ongoing transmission $(z = 0)$, so a scheduling decision needs to be made. We denote the differential cost-to-go function by $S(h,z)$ and the time-average cost by $\lambda$. Then, the Bellman equation is given by:
\begin{multline}
    \label{eq:dynamicProgramming-0}
	S(h,0) = J(\tau,h) + \min_{u\in\{0,1\}}\bigg\{ S(h+1,0), \\ C + S\big(h+1, r(\tau) - 1\big) \bigg\} - \lambda.
\end{multline}
Similarly, we write down the Bellman equation when there is an ongoing transmission. In this case, no scheduling decision needs to be made. When $z>1$, the AoI keeps increasing and the Bellman equation is given by:
\begin{equation}
    \label{eq:dynamicProgramming-1}
	S(h,z) = J(\tau,h) + C + S\big(h+1, z-1 \big) - \lambda.
\end{equation}
When $z = 1$, the AoI drops in the next time-slot and the Bellman recursion is given by:
\begin{equation}
    \label{eq:dynamicProgramming-2}
	S(h,1) = J(\tau,h) + C + S\big(\Delta,0\big) - \lambda.
\end{equation}
Using \eqref{eq:dynamicProgramming-1}, we expand the term $S\big(h+1,r(\tau) - 1\big)$: 
\begin{equation}
    S\big(h+1,r(\tau) - 1\big) = J(\tau,h+1) + C + S\big(h+2, r(\tau) - 2\big) - \lambda.
\end{equation}
Applying \eqref{eq:dynamicProgramming-1} recursively to the right-hand side till we reach $S\big(\Delta, 0)$, we get:
\begin{multline}
    \label{eq:expand_h1}
    S\big(h+1,r(\tau) - 1\big) =  \sum_{k = 1}^{r(\tau)-1}J(\tau,h+k) + C(r(\tau)-1) \\
    + S\big(\Delta, 0\big) - \lambda \big(r(\tau)-1\big).
\end{multline}

Replacing $S\big(h+1,r(\tau) - 1\big)$ in \eqref{eq:dynamicProgramming-0} with \eqref{eq:expand_h1}, we get:
\begin{multline}
    \label{eq:dynamicProgramming-3-u}
	S(h,0) = J(\tau,h) + \min_{u\in\{0,1\}}\bigg\{ S(h+1,0), \\ C r(\tau) + S\big(\Delta, 0 \big) - \lambda \big(r(\tau)-1\big) + \sum_{k = 1}^{r(\tau)-1}J(\tau,h+k) \bigg\} - \lambda.
\end{multline}

Note that now we can simplify the differential cost-to-go function to depend on the AoI only. Let $S'(h) \triangleq S(h,0)$. Then, we get the simplified Bellman equation for our setting: 
\begin{multline}
    \label{eq:dynamicProgramming-3}
	S'(h) = J(\tau,h) + \min_{u\in\{0,1\}}\bigg\{ S'(h+1),  C r(\tau) + S'\big(\Delta\big) \\ - \lambda \big(r(\tau)-1\big) + \sum_{k = 1}^{r(\tau)-1}J(\tau,h+k) \bigg\} - \lambda.
\end{multline}

Without loss of generality, we can set $ S'(\Delta) = 0 $, since $S'(\cdot)$ is a {differential} cost-to-go function.

\myParagraph{Part 1} We consider the case when there exists a threshold $H$ that satisfies the condition \eqref{eq:optimal-threshold}.

We  start by looking at a policy with an arbitrary transmission threshold $ H $, i.e. transmit if and only if the AoI $ h\ge H $. We will show that if $H$ satisfies \eqref{eq:optimal-threshold} then this policy's differential cost-to-go function satisfies the optimal Bellman recursion \eqref{eq:dynamicProgramming-3}. 

To do so, we first compute the differential cost-to-go function for this policy. For all $ h\ge H $, we set $u=1$ in \eqref{eq:dynamicProgramming-3} to get:
\begin{equation}\label{eq:Bellman-dynamics-h-larger-than-threshold}
	\begin{aligned}
		S'(h) &= J(\tau,h) + C\delayComm + \sum_{k=1}^{\delayComm-1}(J(h+k)-\lambda)-\lambda=\\
		&= (C-\lambda)\delayComm + \sum_{k=0}^{\delayComm-1}J(\tau,h+k)
	\end{aligned}
\end{equation}

For $ h = H-1 $ we again use \eqref{eq:dynamicProgramming-3} and set $u=0$ to get:
\begin{equation}\label{eq:Bellman-dynamics-h-H-1}
	\begin{aligned}
		S'(H-1) &= J(\tau,H-1) + S'(H) - \lambda =\\
			   &= \sum_{k=-1}^{\delayComm-1}J(\tau,H+k) - \lambda + (C-\lambda)\delayComm
	\end{aligned}
\end{equation}
where the second equality follows by expanding $S'(H)$ using \eqref{eq:Bellman-dynamics-h-larger-than-threshold}. Repeating this process $j$ times gives us:
\begin{equation}\label{eq:value-function-H-j}
	S'(H-j) = \sum_{k=H-j}^{H+\delayComm-1}J(\tau,k) - j\lambda + (C-\lambda)\delayComm
\end{equation}
Setting $ H-j = \Delta $ in the equation above, we obtain the following equality:
\begin{equation}\label{eq:Bellman-equality-H-1}
	\sum_{k=\Delta}^{H+\delayComm-1}J(\tau,k) - (H-\Delta)\lambda + C\delayComm -\lambda\delayComm = 0.
\end{equation}

Using this, we can compute $ \lambda $:
\begin{equation}\label{eq:lambda-explicit}
	\lambda = \dfrac{\sum\limits_{k=\Delta}^{H+\delayComm-1}J(\tau,k) + C\delayComm}{H+\delayComm-\Delta}.
\end{equation}
For this threshold policy to be optimal,
it has to satisfy the Bellman equation~\eqref{eq:dynamicProgramming-3} such that the minimization procedure over action $u$ computed for each value of AoI $h$ matches the threshold structure.

Thus, for $h = H-1$, the optimal decision must be to not transmit, i.e.
\begin{equation}\label{eq:Bellman-ineq-H-1}
	S'(H) \le C\delayComm + \sum_{k=1}^{\delayComm-1}\bigg(J(\tau,H-1+k)-\lambda\bigg)
\end{equation}
Plugging in the expression of $ S'(H) $ using~\eqref{eq:Bellman-dynamics-h-larger-than-threshold}, we get:
\begin{multline}
    (C-\lambda)\delayComm + \sum_{k=0}^{\delayComm-1}J(\tau,H+k) \le \\ C\delayComm + \sum_{k=1}^{\delayComm-1}\bigg(J(\tau,H-1+k)-\lambda\bigg)
\end{multline}
Simplifying the above yields:
\begin{equation}\label{eq:lambda-low-bound}
	J(\tau,H+\delayComm-1) \le \lambda
\end{equation}
Similarly, for $h = H-2$, we get:
\begin{multline}
    (C-\lambda)\delayComm  - \lambda + \sum_{k=-1}^{\delayComm-1}J(\tau,H+k) \le \\ C\delayComm + \sum_{k=1}^{\delayComm-1}\bigg(J(\tau,H-2+k)-\lambda\bigg).
\end{multline}
Simplifying, we get:
\begin{equation}\label{eq:lambda-low-bound-2}
	J(\tau,H+\delayComm-1) + J(\tau,H+\delayComm-2) \le 2\lambda.
\end{equation}
Repeating the above procedure for any $h<H$, we get:
\begin{equation}\label{eq:lambda-low-bound-3}
	\sum_{k = 1}^{j} J(\tau,H+\delayComm-k) \le j\lambda.
\end{equation}
Observe that due to the monotonicity of the cost function $J(\tau,\cdot)$, the most restrictive of these conditions is \eqref{eq:lambda-low-bound}, since $J(\tau,H+\delayComm-1) \le \lambda$ implies $J(\tau,H+\delayComm-k) \le \lambda, \forall k > 1$ as well. Thus, for it to be optimal to not transmit at any AoI values below the threshold $H$, it is \textit{sufficient} for the following to hold:
\begin{equation}\label{eq:lambda-low-bound-final}
	J(\tau,H+\delayComm-1) \le \lambda
\end{equation}

For AoI $ h = H $, we instead require that the optimal choice be to transmit, i.e. $u = 1$. Thus, the following must hold:
\begin{equation}\label{eq:Bellman-ineq-H}
	C\delayComm + \sum_{k=1}^{\delayComm-1}(J(\tau,H+k)-\lambda) \le S'(H+1)
\end{equation}
Using \eqref{eq:Bellman-dynamics-h-larger-than-threshold} to expand $S(H+1)$, we get:
\begin{multline}
	C\delayComm + \sum_{k=1}^{\delayComm-1}(J(\tau,H+k)-\lambda) \le \\ (C-\lambda)\delayComm + \sum_{k=0}^{\delayComm-1}J(\tau,H+1+k).
\end{multline}
Simplifying the above yields
\begin{equation}\label{eq:lambda-up-bound}
	\lambda \le J(\tau,H+\delayComm).
\end{equation}
Similarly, for $h =  H+1$, we require the optimal decision to be transmit and get:
\begin{multline}
	C\delayComm + \sum_{k=1}^{\delayComm-1}(J(H+1+k)-\lambda) \le S'(H+2) = \\ (C-\lambda)\delayComm + \sum_{k=0}^{\delayComm-1}J(\tau,H+2+k).
\end{multline}
Simplifying the above yields
\begin{equation}\label{eq:lambda-up-boun-2}
	\lambda \le J(\tau,H+\delayComm+1).
\end{equation}
Repeating the above procedure for any value of $h\ge H$, we obtain similar inequalities:
\begin{equation}\label{eq:lambda-up-boun-3}
	\lambda \le J(\tau,h+\delayComm), \forall h\ge H.
\end{equation}
Clearly, the most restrictive of these upper bounds is $\lambda \le J(H+\delayComm)$. Thus, for it to be optimal to transmit at all AoI values $\ge H$, it is sufficient for the following to hold:
\begin{equation}\label{eq:lambda-up-boun-final}
	\lambda \le J(\tau,H+\delayComm).
\end{equation}

The two conditions \eqref{eq:lambda-low-bound-final} and \eqref{eq:lambda-up-boun-final} together imply that if there exists a threshold $H$ that satisfies \eqref{eq:final-condition}, then an optimal policy is to transmit only when the AoI is $\ge H$. 
\begin{equation}
    \label{eq:final-condition}
    J\big(\tau,H+\delayComm-1\big) \le \lambda \le J\big(\tau,H+\delayComm\big).
\end{equation}
Observe that this is identical to the optimal threshold condition \eqref{eq:optimal-threshold} presented in Theorem \ref{thm:optimal-threshold}. This completes one part of the proof.

\myParagraph{Part 2} It still remains to be shown that in case no such threshold can be found, then the optimal policy is to never transmit. For ease of notation, we denote $h+r(\tau)$ as $\widetilde{h}$. Consider the function $V:\mathbb{Z}^{+} \rightarrow \mathbb{R}$, for all AoI values $ h \ge \Delta - r(\tau)+1$, given by:
\begin{equation}
V(h) \triangleq (\widetilde{h}-\Delta)J\big(\tau,\widetilde{h}-1\big) - \sum\limits_{k=\Delta}^{\widetilde{h}-1} J(\delayComp,k), \forall h.
\end{equation}
Observe that for all values of $h \ge \Delta - r(\tau)+1$, we have $V(h+1)-V(h) = (\widetilde{h}+1-\Delta)(J(\tau,\widetilde{h}+1) - J(\tau,\widetilde{h})) \ge 0$. Thus, $V(\cdot)$ is an increasing function. Further, $V(\Delta - r(\tau)+1) = J(\tau,\Delta) - J(\tau,\Delta) = 0$. Thus, $V(h)$ is a non-negative function for all values of AoI $\ge \Delta - r(\tau)+1$. 

Using the function $V(\cdot)$ and the expression for $\lambda$ \eqref{eq:lambda-explicit}, we can rewrite the condition \eqref{eq:final-condition} as follows:
\begin{equation}
    \label{eq:w-condition}
    V({H}) \le C r(\tau) \le  V({H+1}).
\end{equation}

Suppose there exists some $h$ such that $Cr(\tau) \le V({h+1})$. Then, clearly \eqref{eq:w-condition} has a solution at $H = h$, since $V(\cdot)$ is a non-decreasing function. Since we are interested in the case when \eqref{eq:w-condition} does not have a solution, we can safely assume $Cr(\tau) > V({h}), \forall h$. 

Since $V(h) \in [0, Cr(\tau)], \forall h \ge \Delta$, so $V(h)$ converges to a finite value (bounded sequences always converge). The relation $V(h+1)-V(h) = (\widetilde{h}+1-\Delta)(J(\tau,\widetilde{h}+1) - J(\tau,\widetilde{h})) \ge 0$ also ensures that the function $J(\tau,\cdot)$ is bounded. This is because $J(\tau,\widetilde{h})$ is a non-decreasing sequence and has smaller increments than $V(h)$ for each value of $h$. So, we can set $\lambda = \lim_{h \rightarrow \infty} J(\tau, h)$ and $\lambda$ is well-defined.

We also set the differential cost-to-go function to be:
\begin{equation}
    \label{eq:inf-s}
    S'(h) = \sum_{k = h}^{\infty} (J(\tau,k) - \lambda) + Cr(\tau), \forall h \ge \Delta.
\end{equation}
Clearly, $S'(h)$ satisfies the following Bellman recurrence for never transmitting, i.e.
\begin{equation}
    S'(h) = J(\tau,h) + S'(h+1) - \lambda, \forall h \ge \Delta.
\end{equation}
By the monotonicity of $J(\tau,\cdot)$, we know that $J(\tau,h) \le \lambda, \forall h \ge \Delta.$ This, together with \eqref{eq:inf-s} implies 
\begin{equation}
    S'(h+1) \le C r(\tau) + \sum_{k=1}^{r(\tau)-1} (J(\tau,h+k) - \lambda), \forall h \ge \Delta.
\end{equation}
The condition above implies that the minimization procedure to choose $u \in \{0,1\}$ will always select $0$, i.e. never transmit. Thus, our choice of $\lambda$ and $S'(h)$ satisfies the Bellman equations and is optimal. This completes the proof of Theorem 1.


\subsection{Restless Multi-Armed Bandit Formulation}
\label{app:rmab}
We establish that the scheduling optimization described by \eqref{eq:problem-scheduling} is equivalent to a restless multi-armed bandit problem (RMAB).
A restless multi-armed bandit problem \cite{whittle1988restless} consists of $N$ ``arms''. Each arm is a Markov decision process (MDP) with two actions (activate, rest). There are two transition matrices per arm, one describing how the states evolve when the arm is active and one describing how the states evolve when the arm is at rest. Each arm has a cost function mapping states to costs. In the classic RMAB formulation, only one arm can be activated in each time-slot, similar to our scheduling constraint and the goal is to find the schedule that minimizes the long-term time-average cost.

To create a RMAB from \eqref{eq:problem-scheduling}, we first define the arms to represent each agent in the network. The state of every arm $i$ consists of two non-negative integers $(A_i(t), z_i(t))\in\mathbb{Z}^2$. Here, $A_i(t)$ is the AoI of the $i$-th agent while $z_i(t)$ is variable that tracks the number of remaining time-slots to finish an ongoing transmission from agent $i$. Thus, $z_i(t)$ is set to $r_i(\tau_i)-1$ at the start of a new transmission. It decreases by $1$ in each time-slot as the transmission proceeds and is set to $0$ when agent $i$ is not transmitting.

The state evolution of the arm (agent) depends on whether it is currently active (transmitting) or not. If agent $i$ is transmitting in time-slot $t$, then it either initiates a new transmission; or the time remaining to finish sending the current update decreases by $1$. Under this condition, $z_i(t)$ evolves as follows:
\begin{equation}
	(z_i(t+1))_{u_i(t)=1} =
	\begin{cases}
		r_i(\tau_i)-1, &  \text{ if } z_i(t) = 0 \\
		z_i(t)-1, & \text{ otherwise.}
	\end{cases}
\end{equation}
If the agent is transmitting in time-slot $t$ and a new update finished delivery at time-slot $t+1$, i.e. $z_i(t+1) = 0$, then the AoI drops to the age of the delivered packet. Otherwise, the AoI increases by $1$ in every time-slot.
\begin{equation}
	(A_i(t+1))_{u_i(t)=1} =
	\begin{cases}
		\Delta_i, &  \text{ if } z_i(t+1) = 0,\\
		A_i(t)+1, & \text{ otherwise. }
	\end{cases}
\end{equation}

If the agent is not transmitting in time-slot $t$, then there is no update to be delivered and the state evolution is simply given by
\begin{equation}
	\left(A_i(t+1),z_i(t+1)\right)_{u_i(t) = 0} = \left(A_i(t)+1,0\right).
\end{equation}


For every arm $i$, there is a cost function $J_i(\tau_i, A_i(t))$ which maps the state of the arm $(A_i(t),z_i(t))$ to its associated costs, given the processing time allocations $\tau_i$. This completes the MDP specification for each arm.

Since only one arm (agent) can be activated in any time-slot, the goal of the RMAB framework is to find a scheduling policy that minimizes the total time-averaged cost of running the system. Clearly, Markov decision processes evolving as above along with the associated cost functions and activation constraint are equivalent to the scheduling problem~\eqref{eq:problem-scheduling}.

\subsection{Proof of Lemma \ref{lem:indexability}}
\label{app:indexability}
As in Appendix \ref{app:threshold-policy}, we drop the subscript $i$ and establish indexability for a generic agent, since the analysis looks similar for each of the $N$ decoupled problems.

The \textit{indexability} property for the decoupled problem requires that, as the transmission cost $C$ increases from $0$ to $\infty$, the set of AoI values for which it is optimal to transmit must decrease monotonically from the entire set (all ages $A(t)\ge \Delta$) to the empty set (never transmit). In other words, the optimal threshold $H$ should increase as the transmission cost $C$ increases. 

We start with the case when $C = 0$. Clearly, since there is no cost for transmission and the AoI cost function $J(\tau,\cdot)$ is a non-negative increasing function, it is optimal to transmit at every value of AoI $(\forall A(t) \ge \Delta)$.

Let $\widetilde{h} = h+r(\tau)$, as we have used throughout the paper. For $C>0$, we start by defining the function $V:\mathbb{Z}^{+} \rightarrow \mathbb{R}$,  for all AoI values $ h \ge \Delta - r(\tau)+1$, as follows:
\begin{equation}
	V(h) \triangleq (\widetilde{h}-\Delta)J\big(\tau,\widetilde{h} -1\big) - \sum\limits_{k=\Delta}^{\widetilde{h}-1} J(\delayComp,k).
\end{equation}
Observe that for all values of $h \ge \Delta - r(\tau)+1$, we have $V(h+1)-V(h) = (\widetilde{h}+1-\Delta)(J(\tau,\widetilde{h}+1) - J(\tau,\widetilde{h})) \ge 0$. Thus, $V(\cdot)$ is an increasing function. Further, $V(\Delta - r(\tau)+1) = J(\tau,\Delta) - J(\tau,\Delta) = 0$. Thus, $V(h)$ is a non-negative function for all values of AoI $\ge \Delta - r(\tau)+1 $.  

Since $C>0$, there are two possible scenarios - a) there exists $H$ such that $V(H) \leq Cr(\tau) \leq V(H+1)$ or b) $V(h) \leq Cr(\tau), \forall h.$ As proved in Appendix \ref{app:threshold-policy}, if $Cr(\tau) \in \big[V(H),V(H+1)\big)$, then the optimal policy is of threshold type with the threshold being $H$. To map the transmission cost $C$ to a unique optimal threshold, we choose the minimum value of AoI $H$ for which the relation $V(H) \leq Cr(\tau) < V(H+1)$ holds. We call this value $H^*(C)$. When there is no such value of $H$, i.e. $V(h) \leq Cr(\tau), \forall h$ then we set $H^*(C) = \infty$.

Clearly, since the function $V(\cdot)$ is monotone, the optimal threshold $H^*(C)$ is also a non-decreasing function of the transmission cost $C$. This completes the proof of indexability, since we have shown that the set of states for which it is optimal to activate the arm (transmit an update) decreases monotonically as the transmission cost $C$ increases.

The last part of the proof is to derive an expression for the Whittle index. Observe that when $C < V(H+1)/r(\tau)$, the optimal threshold is at $H$ or lower and scheduling decision at $H$ is to always transmit. $C =  V(H+1)/r(\tau) $ is the minimum value of the transmission cost that makes both $H$ and $H+1$ be the optimal threshold, or in other words, makes the transmit and not transmit decisions at AoI $H$ look equally favorable. Thus, the Whittle index is given by:
\begin{equation}
	\begin{aligned}
		W(H) &\triangleq
		\frac{V(H+1)}{r(\tau)} \\
		&=  \frac{\big(\widetilde{H}+1-\Delta\big)J\big(\tau,\widetilde{H}\big) - \sum\limits_{k=\Delta}^{\widetilde{H}} J(\delayComp,k) }{r(\tau)}\\
		&= \frac{\big(\widetilde{H}-\Delta\big)J\big(\tau,\widetilde{H}\big) - \sum\limits_{k=\Delta}^{\widetilde{H}-1} J(\delayComp,k) }{r(\tau)}.
	\end{aligned}
\end{equation}
This completes our derivation of the Whittle index.

\subsection{Entropy Cost as Function of AoI}\label{app:entropy}
In this section, we derive the entropy cost used for the mapping application as a function of the AoI and also establish that it is a monotone increasing function.

Consider the Markov chain describing the occupancy of a cell $c$ in region $i$. Its transition matrix has the following form:
\begin{equation}
	P_i = \begin{bmatrix}
		1-p_i & p_i\\
		p_i & 1-p_i 
	\end{bmatrix}
\end{equation}
The stationary distribution of this Markov chain is $\mu = [0.5, 0.5]$, since $\mu P_i = \mu$. When the base station does not have any update regarding the state of the cell, it sets the probability of occupancy to be $0.5$. The corresponding entropy cost is given by $-\log_2(0.5) = 1$.

Suppose that the base station believes that the cell $c$ is occupied at time $t$ with probability $q$.
At time $t+1$ it does not receive any new update and needs to update its belief about the occupancy of the cell.
Using the transition matrix $P_i$, it updates the distribution to $[q ~~ 1-q] P_i$.
This distribution simply reflects the fact that one time-slot has passed and the base station needs to multiply the original distribution by the state transition matrix to find the current \textit{estimated} state distribution of the cell. This corresponds to the prediction step of a standard Bayes filter.

In fact, this same process is repeated for any general value of AoI. If the last received update about region $i$ says that cell $c$'s state distribution was $[q ~~ 1-q]$, and the current AoI for the region is $A_i$, then the current estimated distribution for cell $c$ at the base station is $\hat{\mu} = [\hat{\mu}_1 ~~ \hat{\mu}_2] \triangleq [q ~~ 1-q] P_i^{A_i}$. The entropy cost for cell $c$ is defined as: 
\begin{equation}
	J_c(A_i) \triangleq -\hat{\mu}_1 \log_2(\hat{\mu}_1) -\hat{\mu}_2 \log_2(\hat{\mu}_2).
\end{equation}
We will show that $J_c(\cdot)$ is an increasing function of the AoI $A_i$, given a fixed value of $q$. Let $\nu_1 \triangleq \hat{\mu}_1(1-p_i) + \hat{\mu}_2 p_i$, and $\nu_2 \triangleq \hat{\mu}_2(1-p_i) + \hat{\mu}_1 p_i$. Then, it is easy to see that:
\begin{equation}
	J_c(A_i+1) = -\nu_1 \log_2(\nu_1) -\nu_2 \log_2(\nu_2).
\end{equation}

Note that the function $x \log_2(x)$ is convex for all $x > 0$, since $\frac{d^2}{dx^2} (x \log_2(x))   = \frac{1}{x} > 0, \forall x > 0.$ Using this fact and the definitions of $\nu_1$ and $\nu_2$, we obtain the following inequalities:
\begin{equation}
    \label{eq:ent1}
    (1-p_i)\hat{\mu}_1 \log_2(\hat{\mu}_1) + p_i \hat{\mu}_2 \log_2(\hat{\mu}_2) \geq \nu_1 \log_2(\nu_1),
\end{equation}
\begin{equation}
    \label{eq:ent2}
    (1-p_i)\hat{\mu}_2 \log_2(\hat{\mu}_2) + p_i \hat{\mu}_1 \log_2(\hat{\mu}_1) \geq \nu_2 \log_2(\nu_2),
\end{equation}

Now, we look at the difference:
\begin{multline}
	J_c(A_i+1) - J_c(A_i) = \\
	\bigg(
	(1-p_i)\hat{\mu}_1 \log_2(\hat{\mu}_1) + p_i \hat{\mu}_2 \log_2(\hat{\mu}_2)   - \nu_1 \log_2(\nu_1)  \bigg) \\ + \bigg(
	p_i\hat{\mu}_1 \log_2(\hat{\mu}_1) +  (1-p_i) \hat{\mu}_2 \log_2(\hat{\mu}_2)  -\nu_2 \log_2(\nu_2)  \bigg)\\
	\ge 0.
\end{multline}
The inequality above follows by applying \eqref{eq:ent1} and \eqref{eq:ent2}. Since $J_c(A_i+1) \ge J_c(A_i), \forall A_i$, so $J_c(\cdot)$ is a monotonic function of the AoI.

While we established this for a single cell in region $i$, the entropy cost of the entire region $J_i(A_i)$ is simply the sum of the entropies of each cell in the region. Thus, the entropy cost functions $J_i(\cdot)$ also grow monotonically with the AoI. 

\begin{figure}
	\centering
	\includegraphics[width=0.97\linewidth]{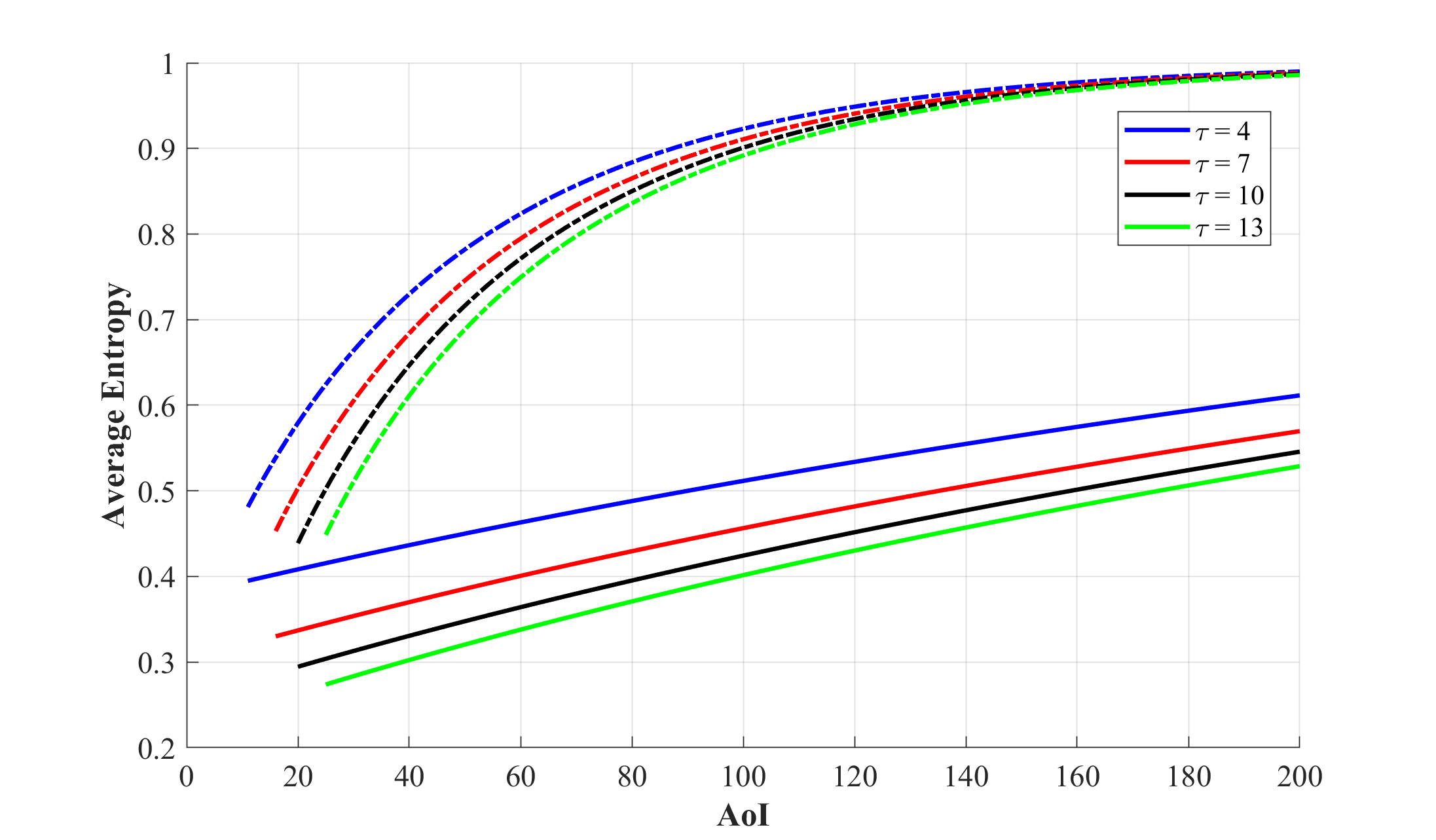}
	\caption{The average entropy of a region v/s AoI. Solid lines represent $p = 0.0005$ and dashed lines represent $p = 0.001$}
	\label{figure:Mapping_cost_function}
\end{figure}

Another point to note is that the probability $q$ reflects the quality of the sent update. If $q$ is close to $0.5$, the update doesn't convey much information about a cell and the entropy cost doesn't drop much on a new update. On the other hand, if $q$ is close to $0$ or $1$, the update contains useful information and the entropy cost drops by a large amount. Since we use sensors that have a limited range and resolutions that improve with the processing time $\tau_i$, the quality of updates also improves for a region with larger $\tau_i$. This ensures that the entropy costs $J_i(\tau_i,A_i)$ satisfy the assumptions required in our co-design framework.

In \autoref{figure:Mapping_cost_function}, we plot the entropy cost as a function of the AoI. We do so by using our sensor for mapping a $40\mathrm{m}\times40\mathrm{m}$ region for different values of processing time $\tau$ and Markov transition probabilities $p$. We observe that the cost grows much more rapidly for the higher value of transition probability $p$. We also observe that for both values of $p$, the entropy cost function starts from a lower value for larger $\tau$, denoting more useful updates for longer processing.

\end{document}